\documentclass[sn-mathphys,Numbered]{sn-jnl}

\usepackage{graphicx}%
\usepackage{multirow}%
\usepackage{amsmath,amssymb,amsfonts}%
\usepackage{amsthm}%
\usepackage{mathrsfs}%
\usepackage[title]{appendix}%
\usepackage{xcolor}%
\usepackage[normalem]{ulem}
\usepackage{textcomp}%

\newif\ifcleanmode
\cleanmodefalse
\ifcleanmode
  \newcommand{\del}[1]{}%
\else
  \newcommand{\del}[1]{\textcolor{red}{\sout{#1}}}%
\fi
\ifcleanmode
  \newcommand{\revonly}[1]{}%
\else
  \newcommand{\revonly}[1]{\textcolor{red}{#1}}%
\fi
\usepackage{manyfoot}%
\usepackage{booktabs}%
\usepackage{algorithm}%
\usepackage{algorithmicx}%
\usepackage{algpseudocode}%
\usepackage{listings}%
\usepackage{geometry}%
\usepackage{url}%

\begin{document}

\title[Neuromorphic Dedispersion for Fast Radio Burst Detection]{Neuromorphic Dedispersion for Fast Radio Burst Detection}

\author*[1]{\fnm{Alessio} \sur{Magro}}\email{alessio.magro@um.edu.mt}

\affil*[1]{\orgdiv{Institute of Space Sciences and Astronomy}, \orgname{University of Malta}, \orgaddress{\city{Msida}, \postcode{MSD 2080}, \country{Malta}}}


\abstract{%
Real-time detection of fast radio bursts at next-generation radio telescopes is increasingly limited by the dedispersion stage, which searches for dispersed pulses and dominates the power and cost of the processing backend.
Neuromorphic processors compute only when input events occur and offer a low-power alternative, but no complete dedispersion pipeline has been demonstrated on them.
Here we present spiking neural dedispersion, a hierarchical delay-and-add pipeline that performs incoherent dedispersion on spike-encoded data, and validate it on synthetic observations of the Northern Cross radio telescope.
Two operating modes trade detection sensitivity against memory and power.
An exact integer-accumulation mode matches the completeness of a standard graphics-processor search at an estimated 68 milliwatts per beam, while a binary mode runs at 1.2 milliwatts and fits entirely on a single neuromorphic chip, retaining most sensitivity for bright, narrow bursts.
Spike-based dedispersion offers a credible low-power backend for wide-field radio-burst surveys.
}%


\maketitle

\section*{Introduction}
\label{introduction}

Fast radio bursts (FRBs) are millisecond-duration radio transients of cosmological origin, first identified by Lorimer et al.~\cite{lorimer2007}, and now established as a heterogeneous population of one-off and repeating sources spanning several orders of magnitude in luminosity~\cite{Petroff2019,petroff2022,cordes2019}. Several thousand individual bursts have been catalogued, with the CHIME/FRB collaboration~\cite{chimefrb2021catalog} providing the largest sample. Their host galaxies span a range of star-formation rates and metallicities~\cite{heintz2020}; the population includes both extragalactic sources and at least one Galactic magnetar~\cite{chimefrb2020magnetar,bochenek2020}; and their dispersion measures sample the warm-hot intergalactic medium with a precision now competitive with conventional cosmological probes~\cite{macquart2020}. The science programme around FRBs has consequently broadened from individual-source astrophysics to cosmological surveys, with applications including independent measurements of the diffuse baryon content of the Universe~\cite{macquart2020,james2022}, dark-energy and Hubble-tension constraints~\cite{hagstotz2022}, and detailed studies of magnetar emission physics~\cite{hessels2019,nimmo2022}. All of these programmes share an operational requirement: continuous, real-time, sensitive single-pulse search across a wide dispersion-measure (DM) range, in commensal mode alongside other observations.

The instrumentation landscape for FRB science is in a period of rapid expansion. Wide-field transit arrays such as CHIME~\cite{chimefrb2018system}, CHORD~\cite{vanderlinde2019chord}, and BURSTT~\cite{lin2022burstt} couple very large fields of view to dense beamforming. Long-baseline arrays such as DSA-2000~\cite{hallinan2019dsa} and the Square Kilometre Array~\cite{braun2015ska} propose continuous wide-field FRB monitoring as a design driver. Existing facilities are upgrading to keep pace: the MeerTRAP backend on MeerKAT~\cite{stappers2017meertrap} and the CRACO upgrade on ASKAP~\cite{wangcraco} substantially expand the dedispersion search volume, typically by factors of several to an order of magnitude in combined beam count, DM trials, and time resolution relative to their direct predecessors. In all of these systems the dedispersion stage is the dominant cost driver of the real-time backend, scaling as $\mathcal{O}(N_b N_f N_{\mathrm{DM}} / \Delta t)$ for $N_b$ beams, $N_f$ frequency channels, $N_{\mathrm{DM}}$ trial DMs, and time resolution $\Delta t$. Operational FRB pipelines in the current generation demand dedicated compute nodes drawing several kilowatts to process groups of simultaneous beams~\cite{sclocco2016}, with aggregate power budgets approaching the megawatt scale on the largest planned systems, making power an increasingly primary constraint in FRB backend design~\cite{pritchard2025rfi}.

The Northern Cross (NC), a T-shaped transit array operated by INAF at the Medicina Radio Observatory (Bologna, Italy), is one of the instruments that has recently undergone a major digital backend upgrade targeting real-time FRB detection~\cite{naldi2017,pellicari2024,naldi2025,naldi2026gass}. Operating at 408\,MHz with 16\,MHz bandwidth and 1024 channels at 0.138\,ms time resolution, it provides a representative operational configuration for a neuromorphic backend: thousands of DM trials, a fine time resolution that stresses the dispersive sweep at high DM, and a transit-mode observing strategy. All numerical results in this paper use the Northern Cross instrument parameters.

Neuromorphic computing performs computation with networks of spiking neurons, taking inspiration from the event-driven, low-power operation of biological nervous systems~\cite{mead1990,schuman2022}. The hardware realising this paradigm spans analogue, mixed-signal and digital designs, both synchronous and asynchronous, unified by massively parallel computation with co-located memory and compute~\cite{schuman2022,davies2021loihi}. The digital, event-driven platforms with low-precision integer arithmetic evaluated here, Intel Loihi~2~\cite{davies2018loihi,davies2021loihi} and SpiNNaker~2~\cite{hoppner2024spinnaker,gonzalez2024spinnaker2}, are among the leading such examples deployed in scientific computing. Compute units are inactive in the absence of input spikes, so energy consumption scales with spike activity rather than with the size of the search problem. Hala Point~\cite{intel_halapoint_2024}, Intel's largest Loihi~2 deployment, supports more than one billion neurons in a 2.6\,kW power envelope, while commercial SpiNNcloud SpiNNaker~2 boards aggregate approximately 7.3 million neurons per 48-chip board. Both architectures support graded-payload spikes, in which a small integer value accompanies the routing event, permitting intermediate computations to propagate partial sums without sacrificing event-driven execution.

Neuromorphic computing in radio astronomy is a recent but rapidly growing field. Pritchard et al.~\cite{pritchard2024rfi} have demonstrated spiking neural network (SNN)-based radio-frequency interference (RFI) detection in radio astronomy data with sensitivity competitive with established methods at substantially lower power. A companion feasibility study~\cite{pritchard2026impact} extends the analysis to FRB detection and 21-cm cosmology on multiple instruments, including the SKA, MWA~\cite{Tingay2013MWA}, ngVLA~\cite{Selina2018ngVLA}, and ASKAP~\cite{Hotan2021ASKAP}, projecting power-consumption ratios of one to two orders of magnitude lower in favour of neuromorphic processing. Deterministic signal-processing operations have previously been implemented on spiking substrates, including fast Fourier transforms~\cite{lopex2022} and graph algorithms~\cite{aimone2019}. The present work moves from the observatory-level feasibility analysis of Pritchard et al.\ to a concrete algorithmic implementation of a complete, validated neuromorphic incoherent dedispersion pipeline. FRB search is well matched to this event-driven model. Astrophysical pulses are rare, so a search pipeline operates predominantly on noise, and a threshold encoder applied to noise produces a sparse spike raster, with typical fire-rate fractions of $10^{-1}$--$10^{-2}$ per channel-sample at detection-calibrated thresholds.

A general FRB search pipeline consists of several stages: incoherent dedispersion across a trial-DM grid, matched filtering of each dedispersed time series against a range of pulse widths, and candidate detection and clustering. The input is \emph{filterbank} data: a stream of detected power in $N_f$ frequency channels, sampled every $\Delta t$. Incoherent dedispersion removes the frequency-dependent dispersive delay by shifting and summing channel power over a grid of trial dispersion measures (DMs). It acts on detected power rather than the raw complex voltages used in coherent dedispersion, and is the standard approach for blind FRB searches over wide DM ranges. The matched filter reduces to a running boxcar sum across multiple widths. This operation can in principle be mapped onto a leaky integrate-and-fire (LIF) neuron with an appropriate decay constant, where threshold crossings serve as detection events; in this work matched filtering is implemented as a classical DM-scaled cumulative-sum operation immediately following dedispersion. Incoherent dedispersion, by contrast, requires careful algorithmic design. A direct implementation costs $\mathcal{O}(N_f)$ operations per trial DM per timestep, and at operational scales requires large neuron and synapse budgets together with a substantial history buffer to hold the dispersive sweep across all channels. Classical fast algorithms such as the Taylor tree~\cite{taylor1974} and the Fast Dispersion Measure Transform (FDMT)~\cite{zackay2017fdmt} reduce the computational cost to $\mathcal{O}(N_f \log N_f)$ per trial DM by sharing partial sums across a hierarchical structure. Direct spike-domain implementations inherit a constraint from each: the Taylor tree is restricted to power-of-two channel pairings and rigid DM grids, and FDMT requires a correction to its standard initialisation when applied to spike-encoded inputs.

The Spiking Neural Dedispersion (SND) algorithm, introduced in this work, belongs to a broader class of event-driven delay-transform algorithms in which a sparse multichannel spike stream is aligned and accumulated over a parameter grid using programmable synaptic delays. Unlike classical delay-and-add architectures, computation occurs only in response to input events, so the energy cost scales with signal activity. SND instantiates this design for the incoherent dedispersion problem, where the parameter grid is the trial-DM space and each per-synapse delay implements the dispersive shift for a channel--DM pair.

We introduce the SND algorithm and a multi-rate extension, a hierarchical delay-and-add tree that performs incoherent dedispersion on spike-encoded data over arbitrary DM grids with a configurable branching factor. We derive the cascade noise statistics that govern its binary mode and characterise the two sensitivity limitations of practical deployment, and we map the full pipeline onto SpiNNaker~2 and Loihi~2 and validate it in software simulation on synthetic Northern Cross filterbanks. An exact integer-accumulation mode reproduces the completeness of a graphics-processor reference search at an estimated 68 milliwatts per beam, while a binary mode fits entirely on a single neuromorphic chip at 1.2 milliwatts and retains most sensitivity for bright, narrow bursts.

\section*{Results}
\label{results}

\begin{figure}[htbp]
  \centering
  \includegraphics[width=0.99\textwidth]{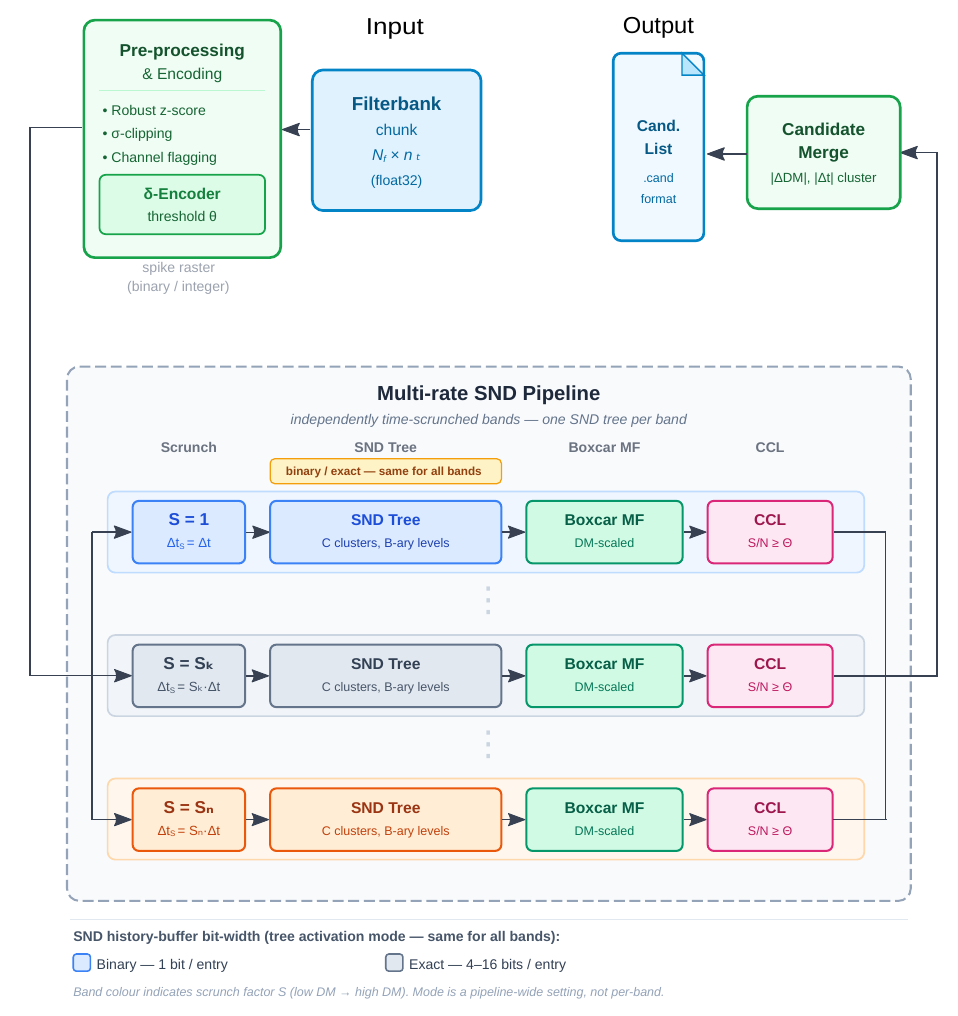}
  \caption{Full architecture of the neuromorphic FRB detection pipeline.
    Raw filterbank chunks are normalised and thresholded by per-channel delta
    encoders to produce a binary spike raster.
    The raster enters the SND hierarchical tree, which produces a DM-time
    output plane for each scrunch band in the multi-rate variant.
    A DM-scaled boxcar matched filter converts each DM-time row into a
    per-cell S/N, and connected-component labelling on the thresholded S/N
    plane extracts candidate events.
    History-buffer bit-widths differ by mode:
    $1$\,bit (binary) and per-level $4$--$16$\,bit (exact).}
  \label{fig:pipeline}
\end{figure}

The complete pipeline, described in detail in Methods, was characterised on synthetic Northern Cross filterbanks: 950 filterbanks containing 12{,}350 injected FRBs that span the full DM and pulse-width range. SND performs incoherent dedispersion as a hierarchical delay-and-add tree over spike-encoded filterbank data. Channels are grouped into small clusters whose per-DM integer delays are summed at leaf nodes, and successive tree levels combine these partial sums with DM-specific delays, forming the DM-time plane in two passes. Two tree-accumulation modes span a trade-off between hardware resources and detection sensitivity: a binary quorum mode that fits entirely on-chip, and an exact integer-accumulation mode that reproduces a full-precision search. Figure~\ref{fig:pipeline} shows the full pipeline architecture. Following the pipeline in order, we report SND performance, cascade noise statistics, hardware resource requirements and power consumption estimates, and validation against the Heimdall reference pipeline~\cite{Barsdell2012Heimdall}, a widely-used GPU single-pulse search package that performs brute-force incoherent dedispersion followed by boxcar matched filtering and candidate clustering. Unless stated otherwise, all numerical results in this section use the Northern Cross instrument parameters ($N_f = 1024$, $\Delta t = 0.138$\,ms, $f_c = 408$\,MHz, 16\,MHz bandwidth, DM $\in [10, 3000]$\,pc\,cm$^{-3}$); a full listing is given in Supplementary Table S1.

\subsection*{SND algorithm}
\label{sec:snd_results}

Practical neuromorphic dedispersion evolves through a sequence of classical approaches, each addressing the limitations of its predecessor. Direct dedispersion scales as $\mathcal{O}(N_f)$ per DM trial and is architecturally straightforward on a spiking substrate, but the total neuron and synapse cost grows prohibitively with the number of DM trials. The Taylor tree~\cite{taylor1974} reduces this to $\mathcal{O}(N_f \log N_f)$ by recursive pair-combination, but imposes a strict binary channel pairing and a power-of-two DM grid. FDMT~\cite{zackay2017fdmt} achieves the same complexity without binary-pairing constraints, but its standard cumulative-sum initialisation was designed for analog radio data and is unsuitable for spike-domain inputs. We observed a sensitivity loss relative to brute-force dedispersion that grows toward low DM, from a factor of about three at high DM to about twelve at the lowest DM tested, which a corrected no-smear initialisation recovers at no additional cost. The structural constraints on the DM grid remain. SND lifts both of these remaining constraints: arbitrary trial-DM grids are supported, the branching factor $B$ is a free parameter matched to the target hardware's synapse-to-neuron ratio, and signal integrity is preserved at every tree level.

\begin{figure}[htbp]
  \centering
  \includegraphics[width=0.9\textwidth]{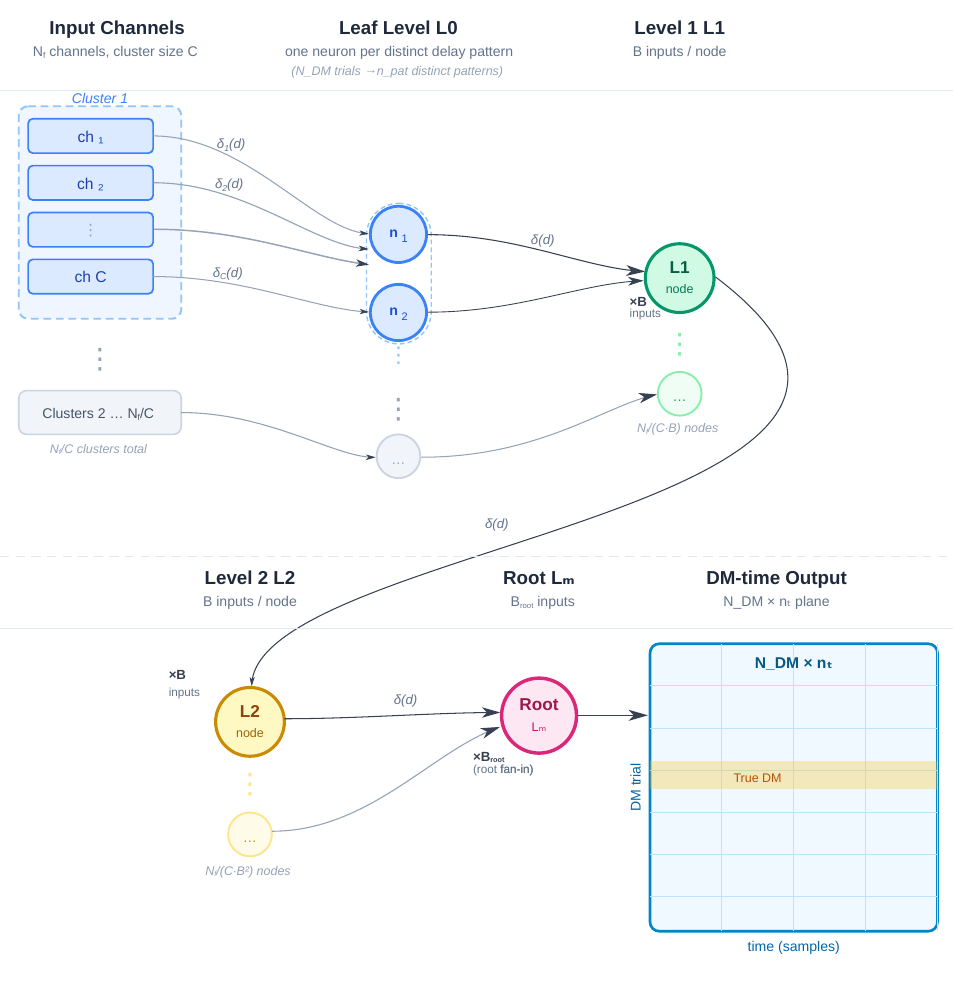}
  \caption{Architecture of the SND dedispersion tree for a single scrunch band.
    Channels are grouped into leaf clusters of size $C$;
    each cluster node accumulates its $C$ channel inputs after applying
    per-synapse delays that implement the intra-cluster dispersive shift.
    Because neighbouring trial DMs share the same integer-delay configuration
    within a narrow channel span, the number of distinct leaf neurons per cluster
    is much smaller than $N_\mathrm{DM}$.
    Above the leaf level, a $B$-ary tree combines cluster
    outputs with per-synapse delays that implement the inter-cluster dispersive shift.
    History buffers (ring buffers) at each level store the delay-line state.}
  \label{fig:snd_arch}
\end{figure}

Figure~\ref{fig:snd_arch} shows the SND architecture. In exact-mode propagation SND recovers detection sensitivity equivalent to brute-force dedispersion across all tested DM and S/N combinations; DM-time planes are numerically invariant under any branching factor $B$, confirming that $B$ is a pure hardware-mapping parameter with no effect on sensitivity. Larger $B$ is preferred on both SpiNNaker~2 and Loihi~2, where synapse-to-neuron capacity ratios of $10^2$--$10^3$ favour wide, shallow trees over narrow, deep ones.

The multi-rate extension partitions the DM range into independently time-scrunched bands. Time scrunching averages groups of $S$ consecutive filterbank samples into one, reducing the time resolution and the per-band data rate by the integer factor $S$, and is applied to the channel power before encoding. At high DM the intra-channel dispersive smearing $\tau_\mathrm{smear}$ is comparable to or larger than the native sample interval, so a pulse of intrinsic width $W_\mathrm{ms}$ spans only $W_S = W_\mathrm{ms} / (S \cdot \Delta t)$ scrunched samples at scrunch factor $S$. Running each DM band at a time resolution matched to $\tau_\mathrm{smear}$ ensures that narrow pulses span at least one scrunched sample, preserving the cascade sensitivity at every DM. A secondary benefit is memory reduction, since scrunching by $S$ reduces the ring-buffer depth at each tree level by the same factor, so the high-DM bands that would otherwise require very deep buffers are kept within a manageable envelope. In binary mode, the 4.6\,MB total history buffer fits within the on-chip SRAM of a single SpiNNaker~2 or Loihi~2 chip, whereas the exact mode (47\,MB) requires external DRAM on either platform.

\subsection*{Spike encoding statistics and cascade noise analysis}
\label{sec:cascade_results}

The sensitivity and false-positive rate of the pipeline depend on the encoder threshold $\theta$ and the tree activation mode. Both modes use a delta encoder, in which each channel is z-score normalised and thresholded. The modes differ only in how the SND tree accumulates the resulting spike values: a 1-bit quorum (binary mode) or exact integer accumulation (exact mode). For Gaussian-normalised inputs the per-channel fire probability under the delta encoder is
\begin{equation}
  p(\theta) = \tfrac{1}{2}\operatorname{erfc}\!\bigl(\theta/\sqrt{2}\bigr).
\end{equation}
At the binary and exact mode operating threshold $\theta = 1.5\sigma$, this yields $p = 6.68 \times 10^{-2}$ theoretically and $7.78 \times 10^{-2}$ measured on Northern Cross data. Raw filterbank power follows a chi-squared distribution with a heavy right tail; a two-pass sigma-clipping normalisation approximately restores Gaussian statistics, recovering this value to within 1--2 percentage points. The measured rate exceeds the theoretical value at $\theta \geq 1.5\sigma$, consistent with the slightly heavier-than-Gaussian tails of the RFI-affected NC noise distribution.

In binary tree mode two thresholds govern the cascade: the \emph{leaf threshold} $K_L$, applied at the cluster level, and the \emph{internal threshold} $K_I$, applied at every level above. A node fires if and only if the number of simultaneously active inputs meets or exceeds its threshold. The noise fire rate propagates as
\begin{align}
  p_1 &= P\!\left(\mathrm{Bin}(C,\, p_0) \geq K_L\right), \nonumber\\
  p_{\ell+1} &= P\!\left(\mathrm{Bin}(B,\, p_\ell) \geq K_I\right), \quad \ell = 1, 2, \ldots \label{eq:cascade}
\end{align}
where $p_0$ is the per-channel encoder fire rate. $K_L$ and $K_I$ are independent parameters: a permissive leaf threshold ($K_L = 1$, any one channel fires the cluster) can be combined with a strict internal threshold, or vice versa. The effective operating constraint is $K_I \leq B_\mathrm{root} - 1 = 3$: the root has $B_\mathrm{root} = 4$ inputs, and $K_I = B_\mathrm{root}$ requires all inputs to fire simultaneously. Integer-delay quantisation systematically distributes signal contributions across adjacent scrunched time slots, making the all-inputs-simultaneous condition unreachable even for arbitrarily bright or wide pulses; $K_I = B_\mathrm{root}$ is therefore excluded. A single $K_I$ is applied at every internal level (L1, L2, and the root); because the root has the smallest fan-in ($B_\mathrm{root} = 4 < B = 8$), it sets the binding constraint. Table~\ref{tab:binaryXX} sweeps the resulting $(K_L, K_I)$ space for the Northern Cross tree, reporting the root-level noise fire rate, the predicted worst-case matched-filter S/N, and the measured S/N at one representative operating point per multi-rate band.

The matched-filter S/N in binary mode exhibits a pulse-width-dependent sensitivity floor. Treating the root output as a Bernoulli variable with signal rate $p_\mathrm{root,sig}$ and noise rate $p_\mathrm{root,noise}$, a boxcar matched filter of width $W$ samples gives
\begin{equation}
  \label{eq:snr_A}
  \mathrm{SNR}_A(W) = \sqrt{W}\;\frac{p_\mathrm{root,sig} - p_\mathrm{root,noise}}{\sqrt{p_\mathrm{root,noise}(1-p_\mathrm{root,noise})}}.
\end{equation}
In noise-suppressing configurations ($p_\mathrm{root,noise} \approx 0$) the denominator is small and the SNR is large; the sensitivity limit is therefore not set by noise propagation. It is set by $p_\mathrm{root,sig} < 1$: for narrow pulses ($W_S = 1$) quantisation scatter distributes channel contributions across adjacent time slots, partially silencing the cascade regardless of signal brightness.

Each cascade level requires a quorum of $K_I$ or more simultaneously active inputs. A pulse of intrinsic width $W_\mathrm{ms}$ spans $W_\mathrm{ms}/(S\!\cdot\!\Delta t)$ scrunched samples; for narrow high-DM FRBs in high-scrunch bands this collapses to $W = 1$ scrunched sample. The entire signal must trigger the cascade in a single time slot. However, the SND tree applies integer-quantised per-channel delays, introducing rounding errors of up to $\pm\tfrac{1}{2}$ native samples per channel. For a $W = 1$ pulse, this scatter distributes channel contributions across two adjacent scrunched samples, with some arriving one slot early and the rest one slot late. At each affected time slot, fewer inputs are simultaneously active, and if the active count falls below $K_I$ at any level the cascade produces zero output for that slot. The result is $p_\mathrm{root,sig} < 1$ even for arbitrarily bright pulses, and the S/N degrades in proportion.

\begin{figure}[htbp]
  \centering
  \includegraphics[width=\textwidth]{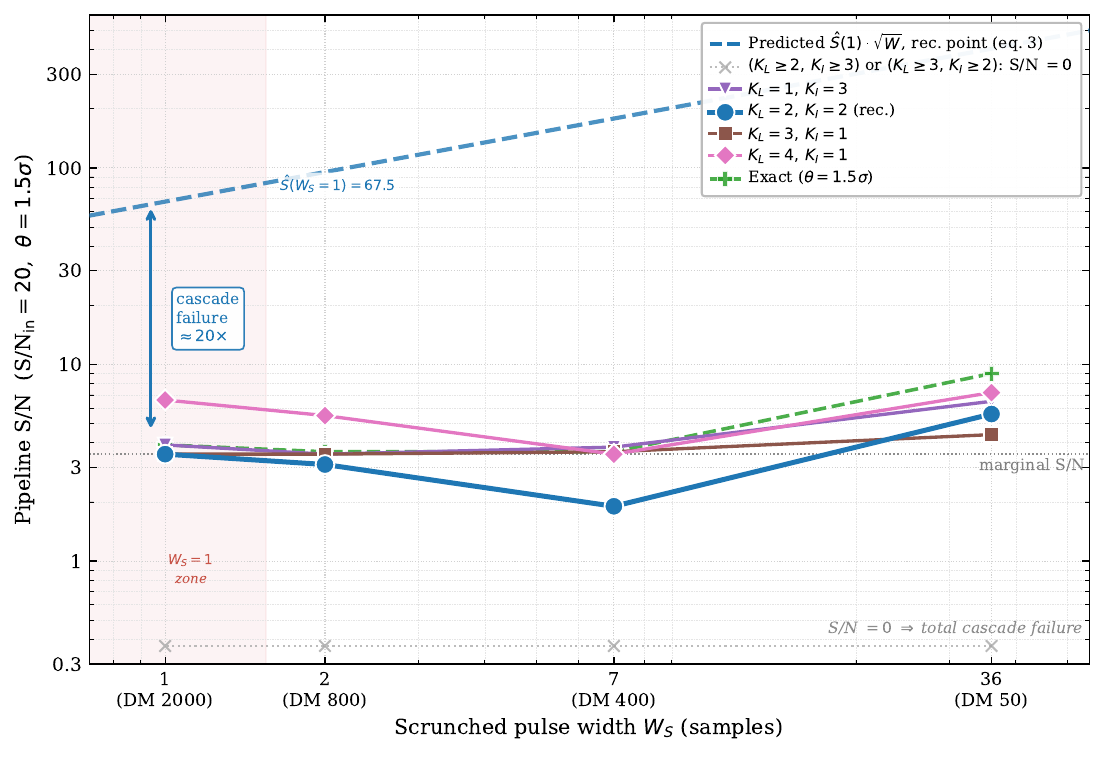}
  \caption{Width-dependent sensitivity floor for binary-mode SND.
    \emph{Dashed curve}: theoretical $\mathrm{SNR}_A(W)$ for the recommended operating point ($K_L{=}2$, $K_I{=}2$, blue).
    \emph{Solid markers}: measured pipeline S/N for a bright injection (S/N$_\mathrm{in}{=}20$) at one representative DM per scrunch band.
    \emph{Grey markers}: cascade over-suppression; measured S/N is zero at all $W_S$.
    For the recommended point, the theoretical prediction at $W_S{=}1$ far exceeds the measured value, revealing cascade failure caused by integer-delay scatter distributing channel contributions across adjacent time slots.
    For $K_I{=}1$ configurations (brown: $K_L{=}3$; pink: $K_L{=}4$), any single branch satisfies the quorum under scatter and measured S/N remains above threshold.
    Reference row (exact, $\theta{=}1.5\sigma$, green) shows no cascade failure.}
  \label{fig:w_floor}
\end{figure}

For wider pulses ($W \geq 2$) the rounding error affects only the leading and trailing edges of the pulse window, and the intermediate samples drive the cascade reliably; $p_\mathrm{root,sig} \approx 1$ and the full $\sqrt{W}$ gain is recovered. Figure~\ref{fig:w_floor} shows the predicted $\mathrm{SNR}_A(W)$ curve from equation~(\ref{eq:snr_A}) alongside the four measured S/N points for each viable $(K_L, K_I)$ configuration. For the recommended point, the $W_S{=}1$ marker is the representative DM\,${\approx}\,2000$\,pc\,cm$^{-3}$ point, and its ${\approx}20\times$ gap below the predicted curve directly visualises the cascade failure.

\begin{figure}[htbp]
  \centering
  \includegraphics[width=0.72\textwidth]{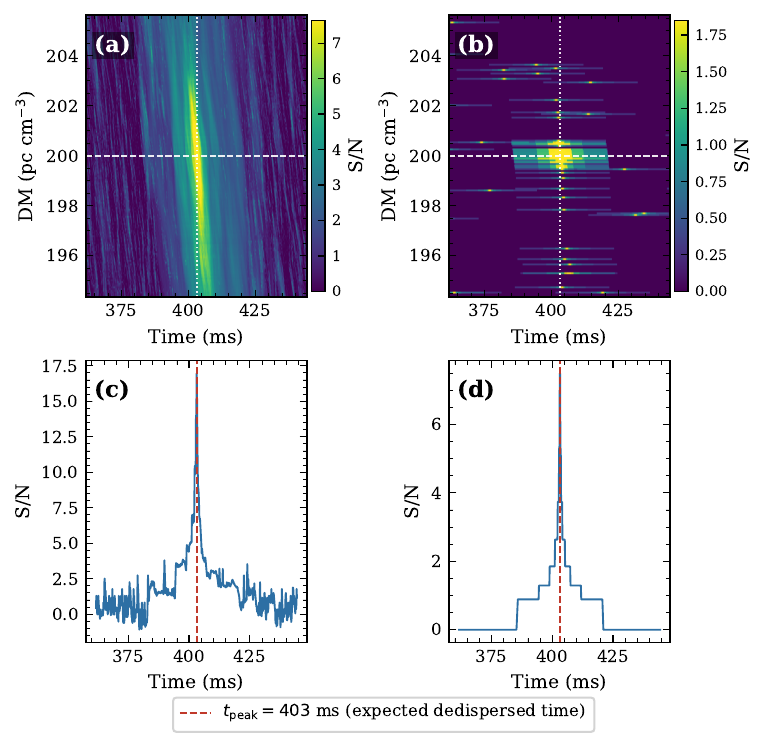}
  \caption{SND tree activation mode comparison on a synthetic bright pulse. The injected pulse has $64$ frequency channels, dispersion measure $200$\,pc\,cm$^{-3}$ and injected signal-to-noise ratio $25$. Panels \textbf{(a)}--\textbf{(b)} show the DM--time matched-filter S/N planes for the exact ($\theta = 1.5\sigma$) and binary ($K_L = 2$, $K_I = 2$, $\theta = 1.5\sigma$) modes respectively; in each plane the dashed white line marks the true DM and the dotted white line marks the expected arrival time. Panels \textbf{(c)}--\textbf{(d)} show the corresponding one-dimensional S/N slice at the true DM for the exact and binary modes, with the dashed red line marking the expected dedispersed time. Both modes recover a localised peak at the correct DM and time. S/N values are in mode-native statistics and are not directly comparable across modes; per-mode completeness is in Table~\ref{tab:mode_summary}.}
  \label{fig:snd_modes}
\end{figure}

In exact mode no quorum is required, so every channel's contribution passes through at every level regardless of timing offsets, and the output is never silenced. The recovered S/N after DM-scaled matched filtering is therefore independent of $W$ in this mode, as follows analytically from the absence of a quorum condition; it serves as the sensitivity ceiling. Figure~\ref{fig:snd_modes} shows DM-time planes and 1-D S/N slices for both modes at a single representative DM and width.

Table~\ref{tab:binaryXX} reports the full sweep. The configuration $K_L = 4$, $K_I = 1$ yields the highest measured S/N for a single bright $W_S = 1$ pulse: with $K_I = 1$ any single branch satisfies the quorum, so quantisation scatter cannot silence the cascade. Configurations $(K_L, K_I) \in \{(2,3),\,(3,2),\,(4,2)\}$ produce zero measured S/N at all $W_S$: the combined leaf and internal quorum is so strict that neither noise nor signal can satisfy it simultaneously, and the cascade is silenced by over-suppression. However, the full-population parameter sweep selects $K_L = 2$, $K_I = 2$ as the recommended operating point: requiring two-out-of-four at the leaf and two-out-of-eight at internal levels provides sufficient noise suppression for the overall false-positive budget while retaining sensitivity to the broader injection population. 

At the recommended $K_L = 2$, $K_I = 2$ point, Table~\ref{tab:binaryXX} shows a measured pipeline S/N of $1.9$ for the single bright test pulse at DM\,400. In the S2 validation population (DM $289$--$578$\,pc\,cm$^{-3}$), intrinsic pulse widths are drawn log-uniformly from $[0.5, 130]$\,ms. At $S = 2$, wider pulses span many more scrunched samples (for example $W_\mathrm{ms} = 16$\,ms gives $W_S = 16/(2 \times 0.138) \approx 58$ samples), and the cascade is reliably triggered over those samples. Those wider events are recovered at high completeness, driving the band total to 47.1\,\%. The cascade failure for $W_S = 1$ pulses in the high-scrunch bands ($S = 8$, $S = 16$) similarly reduces per-band completeness, while wider pulses in the same bands are recovered at close to $100\,\%$ because the multi-sample window provides simultaneous quorum candidates despite rounding. On the full injection population, binary mode reaches approximately 56\,\% aggregate completeness, while exact mode reaches 98.8\,\%, reflecting the sensitivity advantage of unthresholded accumulation.

\begin{table}[htbp]
\centering
\small
\setlength{\tabcolsep}{5pt}
\caption{
  Binary threshold sweep for the Northern Cross SND pipeline.
  Each cell: pipeline S/N for a bright injected pulse (S/N$_\mathrm{in} = 20$) at one representative DM per band:
  DM\,50 ($W_S{=}36$, $S{=}1$), DM\,400 ($W_S{=}7$, $S{=}2$), DM\,800 ($W_S{=}2$, $S{=}4$), DM\,2000 ($W_S{=}1$, $S{=}8$).
  $p_{\rm noise}$: analytic root-level noise fire rate.
  $\hat{S}(W{=}1)$: predicted worst-case S/N (Eq.~\ref{eq:snr_A}).
  \emph{always fires}: $p_{\rm noise} \geq 0.99$; zero S/N: structural cascade failure.
  \textbf{Bold}: recommended operating point.
  Reference row (exact): the sensitivity ceiling in mode-native statistics.}
\label{tab:binaryXX}
\begin{tabular*}{\textwidth}{@{\extracolsep{\fill}} cc rr rrrr @{}}
\toprule
\multirow{2}{*}{$K_L$} & \multirow{2}{*}{$K_I$}
  & \multirow{2}{*}{$p_{\rm noise}$}
  & \multirow{2}{*}{$\hat{S}(W\!=\!1)$}
  & \multicolumn{4}{c}{Measured pipeline S/N} \\
\cmidrule(l){5-8}
& & & & DM\,50 & DM\,400 & DM\,800 & DM\,2000 \\
\midrule
1 & 1 & 1.000            & \multicolumn{1}{c}{---} & \multicolumn{4}{c}{\emph{always fires}} \\
1 & 2 & 1.000            & \multicolumn{1}{c}{---} & \multicolumn{4}{c}{\emph{always fires}} \\
1 & 3 & $2.4\times10^{-1}$ & 1.8  & 6.5 & 3.8 & 3.5 & 3.9 \\
\midrule
2 & 1 & ${\approx}\,1$   & \multicolumn{1}{c}{---} & \multicolumn{4}{c}{\emph{always fires}} \\
2 & 2 & $2.2\times10^{-4}$ & 67.5 & \textbf{5.6} & \textbf{1.9} & \textbf{3.1} & \textbf{3.5} \\
2 & 3 & $<10^{-15}$      & $\infty$  & 0.0 & 0.0 & 0.0 & 0.0 \\
\midrule
3 & 1 & $2.5\times10^{-1}$ & 1.7  & 4.4 & 3.6 & 3.5 & 3.5 \\
3 & 2 & $7.7\times10^{-15}$ & $>10^{7}$ & 0.0 & 0.0 & 0.0 & 0.0 \\
3 & 3 & $<10^{-15}$      & $\infty$  & 0.0 & 0.0 & 0.0 & 0.0 \\
\midrule
4 & 1 & $5.1\times10^{-3}$ & 14.0 & 7.2 & 3.5 & 5.5 & 6.6 \\
4 & 2 & $<10^{-15}$      & $\infty$  & 0.0 & 0.0 & 0.0 & 0.0 \\
4 & 3 & $<10^{-15}$      & $\infty$  & 0.0 & 0.0 & 0.0 & 0.0 \\
\midrule
\multicolumn{2}{@{}l}{exact, $\theta\!=\!1.5\sigma$}  & \multicolumn{1}{c}{---} & \multicolumn{1}{c}{---} & 9.0 & 3.6 & 3.6 & 3.9 \\
\bottomrule
\end{tabular*}
\end{table}

\subsection*{Resource mapping onto neuromorphic hardware}
\label{sec:resources}

Table~\ref{tab:resourcesXX} gives the per-component resource budget for the Northern Cross pipeline. The dominant resource cost is the history ring-buffer, whose size depends on the tree activation mode. The entire pipeline fits on a single neuromorphic chip in terms of neuron and synapse count (144\,000 neurons, 1\,007\,880 synapses); the SpiNNaker~2 chip capacity of $\approx$152\,000 neurons leaves a $5.5\,\%$ margin, but is consistent with published large-network deployments on this platform~\cite{hoppner2024spinnaker}. The binding constraint is therefore memory. In binary mode the 4.6\,MB history buffer fits within the 19\,MB on-chip SRAM of a single SpiNNaker~2 chip, enabling a fully on-chip deployment with one beam per chip. On a 48-chip SpiNNaker~2 board this translates directly to 48 simultaneous beams in binary mode without any external memory. The exact mode exceeds the on-chip SRAM limit and requires DRAM access for the history buffer, whose memory-access power dominates its total budget. The exact mode stores per-level partial sums at their natural integer widths ($4$--$16$\,bit), giving a $47$\,MB footprint at a $5$\,GB/s DRAM bandwidth; naive float32 storage of the same values would require $145$\,MB at $19.5$\,GB/s. At $47$\,MB the exact buffer still exceeds the on-chip SRAM of either chip, so binary mode remains the only fully on-chip configuration.

\begin{table}[htbp]
\centering
\footnotesize
\setlength{\tabcolsep}{3pt}
\caption{Resource budget and chip mapping for the Northern Cross multi-rate SND pipeline.
  \emph{Upper panel}: per-band breakdown; encoder neurons (1024 per band) included in total;
  history buffer given for both tree modes.
  \emph{Lower panel}: chip mapping for the total pipeline (144\,000 neurons, 1\,007\,880 synapses, 6.05\,MB synapse memory).
  Values from the hardware resource model. The exact-mode history buffer stores per-level partial sums at their natural integer widths ($4$--$16$\,bit). Bold indicates the pipeline totals and the chip-fit result.}
\label{tab:resourcesXX}
\begin{tabular}{r r r r r r r}
\toprule
\multirow{2}{*}{$S$} & DM range & \multirow{2}{*}{SND $n$} & \multirow{2}{*}{All $n$} &
  \multirow{2}{*}{Syn.} &
  \multicolumn{2}{c}{Hist.\ buf.\ (MB)} \\
\cmidrule(l){6-7}
 & (pc\,cm$^{-3}$) & & & & Exact & Binary \\
\midrule
 1 &  10--289   & 59\,580 &  60\,604 & 439\,744   & 14.95 & 1.46 \\
 2 & 290--578   & 23\,231 &  24\,255 & 166\,576   &  9.66 & 0.94 \\
 4 & 579--1156  & 23\,475 &  24\,499 & 168\,440   &  9.77 & 0.95 \\
 8 & 1158--2311 & 23\,562 &  24\,586 & 169\,144   &  9.81 & 0.96 \\
16 & 2313--3002 &  9\,032 &  10\,056 &  63\,976   &  2.88 & 0.29 \\
\midrule
\textbf{Total} & & \textbf{138\,880} & \textbf{144\,000} & \textbf{1\,007\,880} &
  \textbf{47.1} & \textbf{4.60} \\
\midrule\midrule
\multicolumn{7}{@{}l}{\textit{Chip mapping}} \\
\midrule
\multicolumn{3}{@{}l}{} & \multicolumn{2}{c}{Binary} & \multicolumn{2}{c}{Exact} \\
\cmidrule(lr){4-5}\cmidrule(l){6-7}
\multicolumn{3}{@{}l}{} & SpiNNaker~2 & Loihi~2 & SpiNNaker~2 & Loihi~2 \\
\midrule
\multicolumn{3}{@{}l}{History buffer} & \multicolumn{2}{c}{4.60\,MB} & \multicolumn{2}{c}{47.1\,MB} \\
\multicolumn{3}{@{}l}{Total memory (hist.+syn.+state)} & \multicolumn{2}{c}{10.8\,MB} & \multicolumn{2}{c}{54.3\,MB} \\
\multicolumn{3}{@{}l}{Fits in chip SRAM?} & \textbf{Yes} & \textbf{Yes} & No & No \\
\bottomrule
\end{tabular}
\end{table}

\subsection*{Power consumption estimate}
\label{sec:power}

All power estimates in this section characterise the SND computation core only, excluding board-level overhead, and represent the dedispersion compute stage in isolation. The power model is evaluated at the recommended operating points for each mode. All figures are model-derived projections computed from published SpiNNaker~2 chip energy parameters~\cite{hoppner2024spinnaker}.

The \emph{static} term, covering neuron leakage and per-timestep SRAM state access, sums to $0.42$\,mW across all five bands and is identical for all tree modes. The \emph{dynamic} term depends on the mean fire rate at each level. In exact mode, the root produces a non-zero output on virtually every timestep (encoder fire rate $7.8\,\%$ per channel, root output rate $96.9\,\%$), giving a dynamic contribution of 4.7\,mW. In binary mode with $K_L{=}2$, $K_I{=}2$, the quorum thresholds suppress most noise events, reducing the measured dynamic contribution to 0.160\,mW.

The \emph{history-buffer} term is the dominant cost in the exact mode, which requires DRAM access at a bandwidth of $5.0$\,GB/s, contributing $62.5$\,mW. Binary mode fits entirely on-chip SRAM, reducing this term to $0.61$\,mW. The history-buffer term applies equally to a Loihi~2 deployment, as it depends only on memory technology.

Grand totals are: exact $\approx$67.6\,mW and binary $\approx$1.2\,mW (from measured fire rates). These projections await hardware validation. Naive float32 storage of the exact mode's partial sums would raise the per-beam power to $248$\,mW at the same completeness. The full per-stage breakdown is in Supplementary Table S2, and a system-level comparison with a GPU reference is given in the Discussion.

The exact mode's integer history buffers are bit-exact with float32 accumulation. Because the delta encoder produces binary spikes, every partial sum in the tree is a bounded non-negative integer that each level stores at its natural width, so the reduced footprint costs no detection sensitivity.

\subsection*{Validation against the Heimdall operational reference}

\begin{figure}[htbp]
  \centering
  \includegraphics[width=\textwidth]{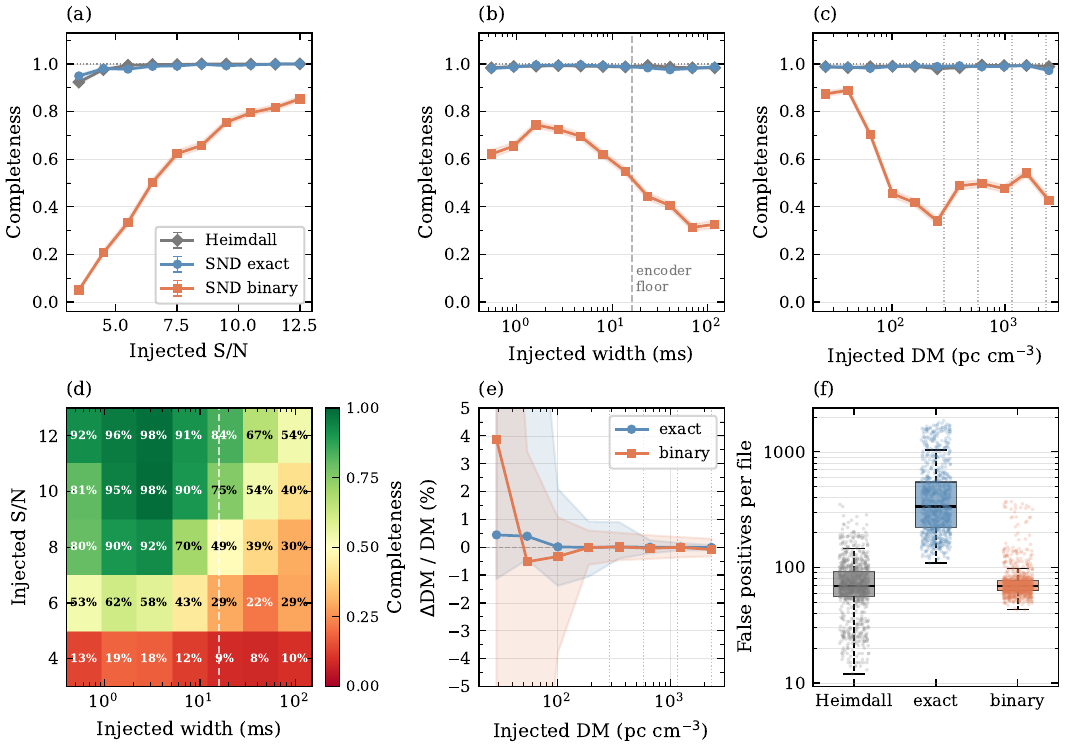}
  \caption{Validation of the spiking neural dedispersion pipeline against the Heimdall reference. Results are from 950 synthetic Northern Cross filterbank observations containing 12{,}350 injected bursts. Pipeline modes are shown as exact (blue circles) and binary (orange squares); the Heimdall reference is shown as grey diamonds.
    \textbf{(a)}~Detection completeness against injected S/N (the injection tool's native normalisation; the recovered matched-filter S/N is systematically higher, Methods).
    \textbf{(b)}~Detection completeness against injected pulse width; the vertical dashed line marks the binary-mode encoder floor ($\approx 16$\,ms in the lowest-scrunch band).
    \textbf{(c)}~Detection completeness against injected DM; the vertical dashed lines mark the scrunch-band boundaries.
    In panels (a)--(c) each point is the detected fraction (completeness); the error bars and shaded bands are the binomial standard error $\sqrt{p(1-p)/n}$, with $n$ the number of injected bursts in that bin ($12{,}350$ in total).
    \textbf{(d)}~Binary-mode completeness as a function of injected S/N and injected width.
    \textbf{(e)}~DM-recovery residual against injected DM; the line is the median residual per bin and the shaded band spans the 16th to 84th percentile, with $n$ the number of matched bursts per bin.
    \textbf{(f)}~False-positive count per file, shown as one marker for each of the $n = 950$ observation files (no error bars).
    The figure uses the relative matching criterion.}
  \label{fig:validation}
\end{figure}

The validation campaign is described in Methods: 950 synthetic Northern Cross filterbanks with a total of 12{,}350 injected FRBs (SGB profile, DM log-uniform in $[20, 3000]$\,pc\,cm$^{-3}$, S/N uniform in $[3, 13]$). The comparison is intended to characterise the sensitivity of the dedispersion stage; SND is run with the components needed for this characterisation: a delta encoder, the multi-rate SND dedispersion tree, per-band DM-scaled matched filters, candidate detection by connected-component labelling, and cross-band candidate merging.

In exact mode, SND detects 12{,}197 of 12{,}350 injections (98.76\,\%), compared to 12{,}211 for Heimdall (98.87\,\%). Among matched candidates, DM residuals have a median of $+0.016$\,\% and an interquartile range of $0.54$\,\%; arrival-time residuals have a median of $-0.14$\,ms and an interquartile range of $1.64$\,ms. Figure~\ref{fig:validation} shows completeness as a function of injected S/N, width, and DM for both modes, and the DM recovery for matched detections.

In binary mode at the recommended operating point, SND achieves a mean per-file detection completeness of 55.7\,\%, compared to Heimdall's 98.9\,\%. Table~\ref{tab:band_completeness} reports completeness by scrunch band for both modes. Table~\ref{tab:mode_summary} summarises the two operating points across all key metrics. The detection threshold reflects a completeness--false-positive trade-off (Supplementary Figure~2): false positives fall steeply as the threshold increases, while binary completeness stays near its maximum up to $\Theta = 7$ and then declines, so $\Theta = 7$ both maximises completeness and keeps false positives well below the lower thresholds. At the matched false-positive rate (${\approx}79$ per file), binary completeness stays below Heimdall; this gap is the cost of the one-bit quorum accumulation and is recovered by the exact mode.

\begin{table}[htbp]
\centering
\small
\caption{Detection completeness by scrunch band on the NC validation set across SND tree activation modes. ``All S/N'' includes all injected $\mathrm{S/N} \in [3,13]$. Exact mode matches Heimdall to within 0.1\% overall (98.8\% vs 98.9\%); both exceed 98.7\% in every band except S16, where exact falls to 96.5\%. Bold indicates the totals over all bands.}
\label{tab:band_completeness}
\begin{tabular}{l r r r r r}
\toprule
Band & DM range (pc\,cm$^{-3}$) & $N_\mathrm{inj}$ & Exact & Binary & Heimdall \\
\midrule
S1  & $[10,\;289]$       & 6602 & 98.8\% & 62.3\% & 98.8\% \\
S2  & $[289,\;578]$      & 1706 & 99.0\% & 47.1\% & 98.8\% \\
S4  & $[578,\;1157]$     & 1678 & 98.7\% & 48.3\% & 98.9\% \\
S8  & $[1157,\;2313]$    & 1757 & 99.0\% & 53.0\% & 99.1\% \\
S16 & $[2313,\;3000]$    & 607  & 96.5\% & 37.2\% & 98.8\% \\
\midrule
\textbf{Total} & & 12{,}350 & \textbf{98.8\%} & \textbf{55.7\%} & \textbf{98.9\%} \\
\bottomrule
\end{tabular}
\end{table}

\begin{table}[htbp]
\centering
\small
\caption{SND pipeline operating-point summary on the NC validation set. Both modes use the delta encoder at $\theta = 1.5\sigma$ and differ only in tree accumulation. Power is the per-beam SND core estimate on SpiNNaker~2 (model-derived from published chip energy parameters); the Heimdall figure is the duty-cycle-adjusted power from a measured single-GPU draw. Completeness is mean per-file fraction of injections matched.}
\label{tab:mode_summary}
\begin{tabular}{l l r r r}
\toprule
Mode & Tree accumulation & Power & Memory & Completeness  \\
\midrule
Binary & 1-bit quorum ($K_L{=}2$, $K_I{=}2$) & 1.2\,mW  &  4.6\,MB & 55.7\%  \\
Exact  & Integer accumulation, $4$--$16$\,bit & 67.6\,mW & 47.1\,MB & 98.8\%  \\
\midrule
Heimdall     & GPU reference                       & 18{,}940\,mW & ---   & 98.9\%  \\
\bottomrule
\end{tabular}
\end{table}

The main sensitivity limitation in binary mode is per-sample encoder under-excitation. For a pulse of intrinsic width $W$\,ms, the per-sample amplitude before dedispersion scales as $W^{-1/2}$. At $\mathrm{S/N} = 7$ and $W = 64$\,ms in the S1 band, the pulse spans $64/0.138 \approx 464$ samples, giving a per-sample flux of $7/\sqrt{464} \approx 0.33\sigma$, well below the $1.5\sigma$ encoder threshold, so effectively no signal enters the tree. For bright ($\mathrm{S/N} \geq 7$), narrow ($W \lesssim 8$\,ms) events, binary SND retains $91\%$ of exact-mode completeness. A secondary DM-dependent structure is also visible. Completeness in the DM $100$--$289$ range, the upper portion of the S1 band, falls to $\sim35\%$, lower than in S2--S8. At these DMs the intra-channel dispersive smearing ($\tau_\mathrm{smear} \approx 0.5$--$2.0$ samples) begins to spread channel contributions across adjacent time slots even at S1 resolution, reducing the coherent addition efficiency of the delta-encoded tree beyond what the cascade and encoder-floor analyses alone predict.

The false-positive rate reflects each mode's noise suppression. Binary SND produces a mean of $78.6$ candidates per file, comparable to Heimdall's $77.1$ . The exact mode yields a mean of $483.8$, as the absence of a quorum threshold allows more noise fluctuations to propagate through the tree. All rates are obtained with only basic RFI mitigation applied: windowed channel flagging via median absolute deviation~\cite{eatough2009} and optional zero-DM subtraction, with no learned classifier. These figures therefore represent a pipeline floor; in practice either post-processing stage would substantially reduce the exact false-positive counts.

\section*{Discussion}
\label{discussion}

\subsection*{Multi-rate architecture and memory}

Without time-scrunching, the history buffer for the full DM range in the exact mode would require 214\,MB per beam, far exceeding the on-chip SRAM of any single SpiNNaker~2 or Loihi~2 chip and making on-chip deployment impossible in either mode. The multi-rate extension partitions the DM search range into bands where each band's time resolution is matched to the intra-channel dispersive smearing at its lowest DM. This brings the exact-mode history buffer down to 47\,MB and the binary-mode buffer to 4.6\,MB, the latter fitting entirely within the on-chip SRAM of a single SpiNNaker~2 chip. The band partition criterion (equation~\ref{eq:scrunch}) ensures that the integer-delay quantisation error at each band boundary remains within the pulse width, so the multi-rate structure introduces no sensitivity loss. As a secondary benefit, the per-band trees in high-DM bands operate at proportionally lower clock rates, reducing dynamic power by the corresponding scrunch factor.

\subsection*{Power-sensitivity trade-off across activation modes}

The two tree activation modes span a trade-off between memory, power, and detection completeness. Binary mode minimises both: a single-chip on-SRAM deployment, an estimated 1.2\,mW per beam, and full compatibility with 48 simultaneous beams on a single SpiNNaker~2 board. The cost is a completeness ceiling of approximately 56\,\% on the mixed-width injection population, set by the two mechanisms analysed above: the $W_S = 1$ cascade failure for narrow pulses in the high-scrunch bands, and the wide-pulse encoder floor above approximately 16\,ms in the S1 band.

Exact mode recovers full sensitivity by accumulating the integer partial sums losslessly, at the cost of a 47\,MB DRAM history buffer per beam and an estimated 67.6\,mW per beam. Removing the quorum condition eliminates the cascade failures that limit binary mode, lifting overall completeness from 55.7\,\% (binary) to 98.8\,\% (exact) and matching the Heimdall reference to within 0.1\%. Wide-pulse sensitivity (the encoder floor) is shared with binary mode and is unchanged.

An intermediate graded mode, which accumulates with an 8-bit cap and no thresholding, was also evaluated. It lifts completeness to 75.3\,\% at 66.8\,mW, but it is Pareto-dominated by exact mode, which reaches 98.8\,\% at essentially the same power and memory. The graded deficit is largest in the high-DM bands (S8 and S16, $\approx$29\,\% and $\approx$9\,\%, with the decline beginning by S4 at $\approx$61\,\%), where the root noise baseline (${\approx}232$ counts) approaches the 8-bit cap of 255 and saturation crushes the dynamic range available to low-S/N signals. Because exact accumulation removes the cap for only a few extra bits per level, graded offers no operating regime of its own, and we do not retain it as a primary mode.

The projected efficiency of binary mode comes mainly from two properties. The 1-bit history buffer shrinks the dominant memory-access cost by $4$--$16\times$ relative to the exact mode, and it resides in on-chip SRAM, which avoids DRAM energy. Both follow from the binary quorum representation, and an FPGA or fixed-function ASIC with the same bit-width and on-chip-memory choices could reproduce them. The event-driven property, that compute units idle when no spikes arrive, reduces only the dynamic term, which is a small part of the binary-mode budget. The power advantage is therefore largely a consequence of a bit-width and memory design that the neuromorphic substrate makes convenient.

Two caveats apply to the comparison with Heimdall. First, the quoted SND figures cover the dedispersion core only; a complete neuromorphic deployment also requires the front-end encoder and the matched-filter stage, which are not event-driven and add a fixed overhead, of order $10^2$\,mW to a few watts per beam for an FPGA or host front-end performing normalisation, encoding and boxcar filtering. Second, the comparison spans both algorithm and implementation. Heimdall performs direct (brute-force) incoherent dedispersion, which scales as $\mathcal{O}(N_f)$ per DM trial~\cite{barsdell2012}, whereas SND is a hierarchical delay-and-add tree in the family of the Taylor tree and FDMT~\cite{taylor1974,zackay2017fdmt} and therefore performs fewer channel-combination operations. The projected power difference is nonetheless dominated by the low-precision, on-chip-memory history buffer rather than by the dedispersion operation count, which contributes only a small fraction of the per-beam budget (Supplementary Table~S2).

\subsection*{Board-level power and deployment}

The per-beam SND power estimates (the power-consumption estimate) characterise the SND core computation only. A realistic deployment requires accounting for the board-level baseline. The SpiNNcloud 48-chip SpiNNaker~2 board draws approximately $100$\,W~\cite{hoppner2024spinnaker}, independent of neural activity. As established in the resource mapping, binary mode fits one beam per chip, giving 48 beams per board.

The marginal power added by SND computation for 48 beams is $\approx$\,57\,mW in binary mode (0.06\,\% of the board baseline) and $\approx$\,3.2\,W in exact mode (3.2\,\%). In both modes the board baseline dominates. Upgrading from binary to exact mode, which recovers the $\approx$\,43\,\% completeness gap, adds 3.2\,W above the $100$\,W board baseline.

The raw filterbank data rate per beam is approximately 14.8\,MB/s (at 16\,bits per value, single polarisation), giving an aggregate input rate of $\approx$\,0.71\,GB/s for 48 beams. This stream is ingested by a front-end processor, which performs normalisation and spike encoding before injection into the neuromorphic system. The resulting spike rate is $\approx 578\,\text{k}$ events/s per beam, or $\approx 27.7$\,M events/s for 48 beams. Although each encoded spike value can be represented using 16\,bits, the transported event stream must also
encode, or otherwise infer from block metadata, the event address and timing. Assuming batched transport and a compact 4--8 byte event representation, this corresponds to $\approx 2.3$--$4.6$\,MB/s per beam, or $\approx 111$--$222$\,MB/s ($\approx 0.89$--$1.78$\,Gbps) for all 48 beams combined, before Ethernet/IP/UDP framing overheads. Batching many events per network packet amortises the transport overhead, but the aggregate rate remains close to, or above, the practical capacity of a single 1\,GbE SpiNNaker\,2 ingress link. The ingress bandwidth would therefore need to be verified experimentally, or mitigated by further event compression, or applying additional upstream rate reduction.

An equivalent GPU deployment for 48 beams requires an estimated 12 RTX~6000~Ada GPUs (four beams per GPU, extrapolated from the measured single-beam throughput), hosted across three four-GPU servers. The single-beam measurement gives an active-window draw of $81$\,W at $23.4\,\%$ duty cycle; at four simultaneous beams the GPU operates near full utilisation ($\approx 93.6\,\%$ duty cycle), drawing power approaching its TDP of approximately 300\,W. With $\approx 250$\,W average sustained draw per GPU card, four GPUs per server consume approximately 1\,kW in GPU power alone; including CPU, memory, networking, and power-supply overhead the realistic total per server is $\gtrsim 1.5$\,kW, giving $\gtrsim 4.5$\,kW for three servers. The 48-beam SpiNNaker~2 board at $100$--$103$\,W therefore suggests a potential system-level power reduction of order $10$--$40\times$ over the GPU cluster in exact mode, depending on GPU occupancy, server infrastructure overhead, and the achievable beams-per-GPU operating point. These system-level estimates compare measured GPU infrastructure against projected SND board power. The per-beam SND compute figures in the power-consumption estimate are algorithmic projections from published chip parameters, and represent order-of-magnitude estimates pending hardware deployment.

\subsection*{Timestep constraints and deployment architecture}

The 0.138\,ms native sampling interval of the Northern Cross is approximately one order of magnitude shorter than the minimum simulation timestep of current neuromorphic chips in large-scale configurations. This reflects a hardware constraint of the current chip generation rather than a fundamental barrier; future chip generations are expected to support sub-millisecond timesteps natively. In the near term, a hybrid front-end can bridge the gap. A dedicated processor, implemented on an FPGA, GPU, or host CPU, performs per-channel sigma-clipping normalisation and delta encoding at the native sample rate, then injects spike events into the neuromorphic fabric via the available board interface (e.g., the SPIF interface~\cite{painkras2012spinnaker} for SpiNNaker~2). The two multi-rate bands with $S \geq 8$ (DM $\gtrsim 1157$\,pc\,cm$^{-3}$) require timesteps of only $\approx 1$\,ms and are already compatible with current SpiNNaker~2 hardware without a dedicated front-end. The finer bands, down to the $0.138$\,ms $S = 1$ band, require correspondingly shorter tree timesteps. On the current chip generation these are reached either by allocating fewer neurons per core, trading aggregate core count for a faster per-core clock, or by folding the finest band into the front-end processor, which already operates at the native sample rate.

As an alternative to a fine-rate front-end, the pipeline could operate directly on hardware pre-averaged data. Averaging by a factor of $S_\mathrm{hw} = 8$ brings the effective sample rate to $1.1$\,ms; for DMs where the intra-channel dispersive smearing satisfies $\tau_\mathrm{smear}(d) \geq S_\mathrm{hw}\cdot\Delta t$, this is consistent with the multi-rate design and introduces no additional sensitivity loss. For lower DMs ($d \lesssim 1157$\,pc\,cm$^{-3}$ for the Northern Cross), however, the coarser rate underresolves the dispersive smearing, degrading sensitivity to narrow FRBs.

The processing chain divides naturally into three layers. A front-end processor handles incoming data, applies per-channel normalisation, and drives the encoder layer. The SND dedispersion tree runs entirely on-chip, producing a sparse stream of DM-time candidates at a rate orders of magnitude below the input. The matched filter and candidate detection operate on this sparse stream and can run on the embedded ARM Cortex-M4F processors on the same SpiNNaker~2 board or be forwarded to a host CPU.

\subsection*{Conclusions}

SND provides a complete, validated incoherent-dedispersion pipeline for neuromorphic hardware, with two activation modes that trade detection sensitivity against memory and power. Exact mode matches the Heimdall reference ($98.8\%$ versus $98.9\%$ completeness) at a projected $68$\,mW per beam, while binary mode reduces the projected per-beam power to $1.2$\,mW at a completeness ceiling near $56\%$, retaining $91\%$ of exact-mode sensitivity for bright, narrow bursts. The power figures are model-derived projections from published chip energy parameters and await hardware validation, but they indicate that spike-based dedispersion is a credible low-power approach for the dedispersion backend of next-generation wide-field FRB surveys.

Richer spike encoding could extend sensitivity beyond the present modes. Both validated modes use the delta encoder, which fires on single-sample excursions above the per-mode threshold. Multi-scale within-band encoding, generating time-averaged copies at $4\times$ and $8\times$ integration within each band before the encoder, would bring sub-threshold wide pulses above the firing threshold at the cost of additional SND trees per band. Alternatively, a multi-threshold encoder can provide graded amplitude information within a binary firing framework. A pre-SND amplification layer, for example a learned matched filter operating on the raw z-scored filterbank before spike conversion, could amplify signals before they enter the tree, shifting the encoder floor.

The candidate-detection stage also admits refinement. Connected-component labelling on the thresholded S/N plane is a simple and robust method but may not be optimal for all pulse morphologies, particularly in the presence of structured RFI or multi-component bursts. Alternative approaches include hierarchical clustering in the (DM, time, width) space, matched-template clustering tuned to the expected dispersive signature, or a lightweight learned classifier operating directly on S/N-plane crops around candidate peaks.

Finally, the power and resource estimates are derived from published chip characterisations. Operational measurements on deployed neuromorphic hardware are needed to validate the projections and quantify system-level overheads, including inter-board communication, host-side preprocessing, and DRAM access patterns under realistic spike traffic.

\section*{Methods}
\label{sec:methods}

\subsection*{Pipeline overview}

The complete pipeline processes a raw filterbank stream in four sequential stages. (1)~\emph{Encoding}: each channel is sigma-clip normalised, z-scored and thresholded to produce a binary or integer-valued spike raster. (2)~\emph{Dedispersion}: the spike raster is passed through the SND tree to produce a DM-time plane. For the multi-rate pipeline, $N_\mathrm{bands}$ independent trees operate on time-scrunched copies of the filterbank in parallel. (3)~\emph{Matched filtering}: a DM-scaled boxcar matched filter computes a per-cell S/N. (4)~\emph{Candidate detection}: connected-component labelling on the thresholded S/N plane extracts candidate events.

All processing is performed in a streaming, chunk-by-chunk fashion. Consecutive chunks overlap by $\delta_\mathrm{max} + 1$ samples, where $\delta_\mathrm{max}$ is the dispersive sweep at the maximum trial DM; this ensures every output sample has a fully loaded delay-line history. Candidates from the overlap region of each non-first chunk are discarded to prevent duplicates.

\subsection*{Neuromorphic hardware platforms}
\label{sec:hardware}

The pipeline is mapped onto two contemporary neuromorphic platforms: SpiNNaker~2 and Loihi~2. Key specifications are summarised in Supplementary Table S3.

\textbf{SpiNNaker~2}~\cite{mayr2019spinnaker,hoppner2024spinnaker,gonzalez2024spinnaker2} was developed at TU~Dresden and the University of Manchester. Each chip integrates 152 ARM Cortex-M4F processing elements (PEs), each with 128\,KB of private SRAM, for a total of approximately 19\,MB of on-chip SRAM per die. Spike communication is packet-based, with each event carrying a routing key and a 32-bit payload that enables graded-value propagation through the tree. Axonal delays are supported at single-timestep resolution via the routing schedule, directly implementing the dispersive shift $\delta_c(d)$ without additional circuitry. Per-spike energy consumption on this process node is estimated at $E_\mathrm{spike} \approx 10$\,pJ~\cite{hoppner2024spinnaker}. The SpiNNcloud commercial board aggregates 48 chips at a board-level power of approximately 100\,W, supporting approximately 7.3 million neurons per board at typical software neuron densities.

\textbf{Loihi~2}~\cite{davies2021loihi}~\cite{intel_loihi2_brief} is Intel's second-generation neuromorphic research chip. Each chip contains 128 neuromorphic cores, each supporting up to 8192 neuron compartments, giving a total capacity of approximately 1\,million neurons per chip. Per-synapse delays are configurable in integer timestep increments up to 63 timesteps; this limit applies to direct synaptic connections between adjacent tree nodes, not to the full dispersive sweep. In SND the long-range delays are implemented as ring-buffer read offsets in neuron compartment SRAM, so the 63-timestep constraint does not cap the maximum processable DM. Spike messages carry a programmable integer payload of up to 32\,bits; the cores are integer-only, which directly suits the exact mode's integer accumulation. The chip draws approximately 1\,W in typical operating configurations at a 1\,ms timestep; a per-synaptic-operation energy of ${\approx}10$--$12$\,pJ has been reported for a sensor-fusion benchmark~\cite{isik2024loihi2}, though a direct per-spike figure analogous to SpiNNaker~2's packet-energy measurement is not available. The Hala Point system~\cite{intel_halapoint_2024}, aggregates 1152 Loihi~2 chips within a 2.6\,kW power envelope to provide more than one billion neurons. Unlike SpiNNaker~2's single shared SRAM pool, Loihi~2 SRAM is distributed per neurocore, with each of the 128 cores on a chip holding approximately 192\,KB of local compartment and synapse RAM, giving a total of roughly 24\,MB per chip. In binary mode the 4.6\,MB history buffer fits within this on-chip budget when distributed across cores; the exact mode (47\,MB) exceeds it and requires history-buffer DRAM access on Loihi~2, as on SpiNNaker~2.

Both platforms provide the three hardware capabilities on which the SND mapping relies. (i)~\emph{Per-synapse programmable integer delays}: each synaptic connection carries a delay that offsets the spike's effective arrival time by the corresponding number of timesteps, natively realising the dispersive shift $\delta_c(d)$ for each channel and trial DM. (ii)~\emph{Multicast routing}: a single neuron's output spike is delivered simultaneously to an arbitrary set of downstream subscribers, so each channel encoder drives all DM-trial neurons at the appropriate delay without explicit packet replication. (iii)~\emph{Graded payloads}: integer partial sums are propagated between tree levels in exact mode, preserving accumulation precision through the full tree depth.

Quantitative power figures in this work use SpiNNaker~2 energy parameters throughout. Extending them to Loihi~2 is only a first-order approximation. The two platforms differ in process node, memory hierarchy and routing, so only the history-buffer term, which is set by the memory technology, carries over directly. A platform-specific Loihi~2 estimate would require a per-spike energy figure comparable to SpiNNaker~2's, and only a benchmark-specific per-synaptic-operation value is currently available for Loihi~2.

\subsection*{The neuromorphic dedispersion neuron}

The dedispersion problem at trial DM $d$ is to compute the channel-aligned sum
\begin{equation}
  y_d(t) = \sum_{c=1}^{N_f} x_c\bigl(t - \delta_c(d)\bigr),
\end{equation}
where $x_c(t)$ is the channel-$c$ input and $\delta_c(d)$ is the integer dispersive delay. In the spike-based setting, the continuous channel power $x_c(t)$ is replaced by a binary or integer-valued spike stream produced by a channel-level encoder. Each summing operation in the algorithm is implemented as an integrate-and-fire neuron with no leak term, implementing a plain delayed sum (Eq.~\ref{eq:snd_neuron}) whose input synapses carry programmable per-connection delays:
\begin{equation}
  v(t) = \sum_{i} w_i \cdot s_i\bigl(t - \delta_i\bigr),
  \label{eq:snd_neuron}
\end{equation}
with $s_i(\cdot)$ the spike train from input $i$ and synaptic weight $w_i$ (unity for dedispersion). The neuron emits an output spike when $v(t)$ crosses a configurable threshold; in exact-mode propagation the partial sum is forwarded directly without thresholding, preserving exact arithmetic.

\subsection*{Input encoding}
\label{sec:encoding}

The encoding layer converts the raw per-channel power spectrogram $P_c(t)$ into a spike raster $x_c(t)$ before entering the SND tree. Raw filterbank power follows an approximately exponential (chi-squared) distribution. A two-pass sigma-clipping normalisation is applied before encoding: (1) compute initial $\mu_0, \sigma_0$ per channel; (2) clip $P_c$ to $[\mu_0 - 5\sigma_0, \mu_0 + 5\sigma_0]$; (3) recompute $\mu, \sigma$ from the clipped values; (4) apply clipped statistics to the original data. This recovers the theoretical fire rates quoted in the cascade noise analysis to within $\sim$2--3 percentage points across the encoder thresholds on real Northern Cross data. Both pipeline modes use a delta encoder in which each channel is z-score normalised and thresholded,
\begin{equation}
  x_c(t) = \mathbf{1}\!\left[z_c(t) > \theta\right], \qquad z_c(t) = \frac{P_c(t) - \mu_c}{\sigma_c}.
\end{equation}

For zero-mean Gaussian $z$, the per-sample fire probability is $p(\theta) = \tfrac{1}{2}\operatorname{erfc}(\theta/\sqrt{2})$. The per-channel signal is individually sub-threshold, yet this does not preclude detection. An injection of matched-filter S/N $\rho$ distributed over $N_f$ channels and $W$ samples has a single channel-sample peak amplitude of $\rho/\sqrt{N_f W}$, which for $N_f = 1024$ lies well below the $\theta = 1.5\sigma$ delta-encoder threshold for every injected event in the validation set. Detection does not require the signal to cross the threshold in any individual channel. The signal instead raises the firing probability of each signal channel-sample by $\delta p \approx (\rho/\sqrt{N_f W})\,\varphi(\theta)$, where $\varphi$ is the standard normal density. Summed along the dispersion track, the excess spike count is $N_f W\,\delta p = \sqrt{N_f W}\,\rho\,\varphi(\theta)$; measured against the $\sqrt{N_f W\,p(\theta)\,[1 - p(\theta)]}$ counting noise, this gives a detection S/N of $\rho\,\varphi(\theta)/\sqrt{p(\theta)\,[1 - p(\theta)]}$, which is independent of $N_f$ and $W$ and proportional to $\rho$. The tree accumulation and matched filter therefore recover the injected S/N up to a fixed efficiency factor ($\approx 0.5$ at $\theta = 1.5\sigma$). This fixed quantisation loss sets the binary-mode sensitivity.

\subsection*{Incoherent dedispersion}

The cold-plasma dispersive delay between two channels at frequencies $f_i$ and $f_j$ (in MHz) at dispersion measure DM (in pc\,cm$^{-3}$) is
\begin{equation}
  \Delta t_{ij}(\mathrm{DM}) = K_{\mathrm{DM}} \cdot \mathrm{DM} \cdot (f_i^{-2} - f_j^{-2}),
\end{equation}
with $K_{\mathrm{DM}} \approx 4.149 \times 10^6$\,MHz$^2$\,ms\,pc$^{-1}$\,cm$^3$~\cite{lorimer2004handbook}. Incoherent dedispersion at trial DM $d$ produces the channel-aligned time series $y_d(t) = \sum_{c=1}^{N_f} P_c(t + \delta_c(d))$, where $P_c(t)$ is the power in channel $c$ and $\delta_c(d) = \lfloor \Delta t_{c, c_{\mathrm{ref}}}(d) / \Delta t \rceil$ is the integer per-channel sample delay relative to a reference channel. A burst is detected by applying a multi-width matched filter to $y_d(t)$ and thresholding the result. Trial DMs are spaced on a quasi-geometric grid following Heimdall, with adjacent trial separation set by the DM step at which the total dispersive broadening across the band remains within a tolerance factor $\tau$ of the intrinsic pulse width; we use $\tau = 1.05$ throughout to match the Heimdall configuration.

\subsection*{Classical dedispersion algorithms}
\label{sec:taylor}

The Taylor tree~\cite{taylor1974} operates bottom-up by recursive pair-combination of adjacent channels. After $\log_2 N_f$ levels the root produces one dedispersed time series per trial DM. Its structural constraint is that it operates on a strict binary pairing of adjacent channels, and the natively supported trial DMs are those whose integer pair-delays form a base-2 ladder; arbitrary DM grids require re-gridding at the leaves.

\label{sec:fdmt}
FDMT~\cite{zackay2017fdmt} avoids the binary-pairing constraint by merging adjacent sub-bands iteratively, computing the dispersive shift across each sub-band from the $f^{-2}$ scaling law. After $\log_2 N_f$ iterations the root produces one dedispersed time series per output $\delta$. The standard FDMT initialisation sets $\mathrm{slab}[c, \delta, t] = \sum_{k=0}^{\delta} \mathrm{input}[c, t+k]$ (cumulative sum), acting as a matched filter for within-channel dispersive smearing in analog radio data. When applied to spike-encoded inputs, which carry no intra-channel smearing in the analog sense, this cumulative-sum step redistributes each discrete spike across multiple $\delta$-bins, reducing peak sensitivity relative to brute-force dedispersion by a factor that grows toward low DM, from ${\approx}3\times$ at high DM to ${\approx}12\times$ at the lowest DM tested (Supplementary Figure~1). The no-smear initialisation sets $\mathrm{slab}[c, \delta, t] = \mathrm{input}[c, t]$ for all $\delta$; the downstream recursion is unchanged and most of the sensitivity is recovered. SND routes each spike to a single $(d, t)$ cell and matches brute-force dedispersion exactly (Supplementary Figure~1). Neither the Taylor tree nor the corrected FDMT supports fully arbitrary trial-DM grids or configurable branching factors matched to the hardware's synapse-to-neuron ratio; these are the motivations for SND.

\subsection*{The Spiking Neural Dedispersion algorithm}
\label{sec:snd}

The Taylor and FDMT ports establish feasibility for spike-domain dedispersion, but neither fully exploits the freedoms of the neuromorphic substrate. On a GPU, the deep memory hierarchy and the cost of irregular memory access shape the algorithm. Channel pairing must be regular to enable coalesced reads, and DM grids must be structured to enable shared partial-sum reuse. Neuromorphic hardware imposes neither constraint. Per-synapse programmable delays mean that any integer time alignment is equally cheap. Multicast routing means that a single neuron's output can be delivered to an arbitrary set of downstream subscribers. There is therefore no architectural reason to restrict the dedispersion tree to a binary pairing of adjacent channels or to require a power-of-two DM grid.

SND accepts an arbitrary trial-DM grid as input and constructs a dedispersion plan in three stages.

\subsubsection*{Leaf clusters}
Channels are partitioned into contiguous clusters of size $C$. Within each cluster, the per-channel dispersive delay $\delta_c(d) - \delta_{c_0}(d)$ relative to the cluster's reference channel $c_0$ is computed for each trial DM $d$. The distinct integer-delay patterns within a cluster are enumerated. Each distinct pattern becomes one leaf neuron, which sums the $C$ delay-aligned channel inputs. For typical configurations the number of distinct patterns per cluster is much smaller than $N_{\mathrm{DM}}$ because nearby trial DMs share the same integer-delay pattern at small channel spans.

\subsubsection*{Tree levels}
Above the leaves, a $B$-ary tree combines cluster outputs with inter-cluster delay alignment for each trial DM. At each level, a node combines the outputs of $B$ children spanning adjacent sub-bands. The per-child delay is the integer inter-band delay for that trial DM. The total number of internal levels is $L = \lceil \log_B (N_f / C) \rceil$. Distinct sub-DM patterns at each internal node are enumerated as at the leaves, each becoming one neuron at that node.

\subsubsection*{Tree activation modes}
Two activation modes govern how each node combines its $B$ child inputs and how the root output is normalised before the matched filter; both use the delta encoder and differ only in tree accumulation precision (full operating points are listed in Supplementary Table S1).
\begin{itemize}
  \item \textbf{Binary} (configurable per-level quorum thresholds $K_L$, $K_I$): each node emits $\{0,1\}$ by thresholding its accumulated input count. History buffers require 1\,bit per entry. The viable operating window and sensitivity floor are analysed in the cascade noise analysis.
    \item \textbf{Exact} (integer accumulation, no thresholding): each node forwards the exact integer sum of its inputs at its natural bit width. The root output is divided by $\sqrt{N_f}$ to produce per-channel-equivalent units. History buffers store each level's partial sums at their natural integer width ($4$--$16$\,bit); float32 storage is bit-identical but requires 32\,bits per entry.
\end{itemize}

In exact-mode propagation SND is mathematically invariant under the choice of $B$; DM-time planes are numerically identical for $B \in \{2, 4, 8, 16\}$ on the same input. Increasing $B$ reduces tree depth and total neuron count while increasing per-node fan-in; both SpiNNaker~2 and Loihi~2 maintain synapse-to-neuron ratios in the range $10^2$--$10^3$, so larger $B$ is generally preferred.

\subsection*{Multi-rate extension}
\label{sec:multirate}

Time scrunching is applied to the raw channel power, before encoding. Each band averages $S$ consecutive native samples of the filterbank into one scrunched sample, and the per-channel delta encoder is then applied independently within each band at that band's scrunched resolution. The binary/exact distinction therefore arises only in the tree accumulation.

Running separate DM sub-bands at different time resolutions is established practice in GPU single-pulse search, for example the DDplan sub-band scheme~\cite{ransom2001presto} and Heimdall's adaptive time-scrunching. The multi-rate SND pipeline adapts this to the neuromorphic setting. The element specific to SND is that each tree node carries its own integer delay pattern, so per-band scrunching is realised directly through per-synapse delays, and the time series is not re-binned between bands.

The DM range is partitioned into bands $(d_{\ell}, d_{\ell+1})$ by the criterion that the scrunch factor $S$ at the lower band edge satisfies
\begin{equation}
  \label{eq:scrunch}
  S \cdot \Delta t \;\leq\; \tfrac{1}{2}\,\tau_{\mathrm{smear}}(d_\ell),
  \qquad
  \tau_{\mathrm{smear}}(d_\ell) = 2\,K_{\mathrm{DM}}\, d_\ell \,\frac{\Delta f}{f_c^3},
\end{equation}
where $\Delta f$ is the per-channel bandwidth and $f_c$ is the centre frequency. Here $\tau_\mathrm{smear}$ is the intra-channel dispersive smearing at the band's lowest DM, the maximum broadening of a delta-function pulse within a single frequency channel. The criterion keeps the scrunched sample interval within half this width, a conservative margin, so time-averaging by factor $S$ removes only information already blurred by dispersion. Equivalently, the SND integer-delay quantisation error ($\pm S/2$ native samples), which scales with $S$, remains within the pulse width at each band boundary. For the Northern Cross (Supplementary Table S1) this criterion produces five bands: $S = 1$ for $d \lesssim 289$\,pc\,cm$^{-3}$, $S = 2$ for $289 \lesssim d \lesssim 578$\,pc\,cm$^{-3}$, $S = 4$ for $578 \lesssim d \lesssim 1157$\,pc\,cm$^{-3}$, $S = 8$ for $1157 \lesssim d \lesssim 2313$\,pc\,cm$^{-3}$, and $S = 16$ for $d \gtrsim 2313$\,pc\,cm$^{-3}$. Each band runs an independent SND tree with an instrument descriptor using $S \cdot \Delta t$ as its effective sample interval.

After per-band candidate detection, candidates from all bands are merged by clustering in (DM, time): two candidates are merged if $|\Delta \mathrm{DM}| \leq 20$\,pc\,cm$^{-3}$ and $|\Delta t| \leq 200$\,ms, retaining the highest-S/N representative.

\subsection*{Matched filtering}
\label{sec:mf}

A DM-scaled boxcar matched filter~\cite{cordes2003transients} is applied to each row of the DM-time plane. The set of boxcar widths at row $k$ is
\begin{equation}
  \mathcal{W}(d_k) = \mathcal{W}_\mathrm{base} \;\cup\; \left\{w \in \{16, 32, 64, \ldots, 1024\} : w \leq \alpha \cdot \tau_\mathrm{sweep}(d_k)/\Delta t\right\},
\end{equation}
where $\mathcal{W}_\mathrm{base} = \{1, 2, 4, 8\}$, $\tau_\mathrm{sweep}(d_k) = K_\mathrm{DM}\,d_k\,(f_\mathrm{lo}^{-2} - f_\mathrm{hi}^{-2})$ is the total dispersive sweep time across the band at trial DM $d_k$, and $\alpha = 0.1$ compensates for accumulated integer-delay rounding in the tree; this caps the longest boxcar at one-tenth of the band's dispersive sweep time, keeping the matched-filter window within the timescale over which integer-delay rounding would otherwise smear the coherent sum. Each boxcar is evaluated in $O(n_t)$ time using a precomputed cumulative-sum array. The per-cell S/N is
\begin{equation}
  \mathrm{SNR}(k, t, W) = \frac{\sum_{t'=t}^{t+W-1} y_{d_k}(t') - \mu_k\, W}{\sigma_k \sqrt{W}},
\end{equation}
where $\mu_k$ and $\sigma_k$ are estimated per row from the mean and standard deviation of valid samples. For binary planes, where the natural per-row $\sigma_k \sim \sqrt{p_\mathrm{root}(1 - p_\mathrm{root})} \ll 1$, a noise floor $\sigma_\mathrm{min} = {}$0.20 is applied to prevent isolated single-fire events from producing artificially large S/N: with this floor a single fire in a $W = 64$ boxcar yields $\mathrm{SNR} \approx {}$0.6, while a cluster of $\sim$25 fires in the same window yields $\mathrm{SNR} \approx {}$16.

\subsection*{Candidate detection}
\label{sec:candidates}

Candidate events are extracted from the S/N plane by: (1)~thresholding at a configurable detection S/N $\Theta$ (default 6.0--10.0); (2)~applying a binary closing operation with a $9 \times 9$ structuring element to bridge single-pixel gaps between adjacent peaks; (3)~running connected-component labelling (CCL) on the closed mask using 8-connectivity; (4)~reporting one candidate per connected island, with S/N and (DM, $t$) taken from the peak pixel. Islands with fewer than $N_\mathrm{min}$ pixels (default 10) are discarded to suppress isolated noise spikes that pass the S/N threshold in sparse binary planes.

This approach differs from the standard Heimdall clustering, which groups candidates by proximity in the (DM, time) plane and selects the peak. The CCL approach used here is well-suited to the SND output because binary planes can produce sparse, multi-pixel islands with irregular shapes; the closing operation consolidates them before labelling. The small-island filter ($N_\mathrm{min}$) replaces Heimdall's member-count threshold. The equivalence of CCL and Heimdall's clustering on exact-mode SND output is confirmed by the near-identical detection completeness of exact SND (98.8\,\%) and Heimdall (98.9\,\%) reported in Table~\ref{tab:mode_summary}. The matched-candidate DM and arrival-time residuals also agree closely, consistent with the two clustering schemes identifying the same candidates.

\subsection*{Hardware mapping and power modelling}

For a pipelined SND implementation, each level-$\ell$ neuron must hold a ring buffer of past
output values so that level-$(\ell+1)$ parents can read them at an integer time offset.
Let $d_{n,m}$ denote the minimum ring-buffer depth (in scrunched samples) for node $n$ at
DM configuration $m$; this is the maximum shift any parent node requests from node $n$,
precomputed during plan construction.  Input channels additionally require a spike ring
buffer of depth $d_\mathrm{in}$, the maximum intra-cluster delay across all DM configurations.
The total history-buffer memory for one scrunch band is
\begin{equation}
  \label{eq:hist_mem}
  M_\mathrm{band} = \left\lceil \frac{N_f \cdot d_\mathrm{in} \cdot b_\mathrm{in}}{8} \right\rceil
    + \sum_{\ell=0}^{L-1} \sum_{n} \sum_{m}
      \left\lceil \frac{d_{n,m} \cdot b_\mathrm{out}}{8} \right\rceil,
\end{equation}
where $b_\mathrm{in} = 1$\,bit/sample (encoder output is always binary), $b_\mathrm{out}$ is
the inter-level bit width ($1$\,bit for binary, per-level $4$--$16$\,bit for exact),
and the ceiling is applied \emph{per DM configuration} $m$ rather than once over the aggregate:
each DM configuration has its own ring-buffer pointer and size, so allocations cannot share
padding bytes across configurations.
The total across all bands is $M_\mathrm{hist} = \sum_\mathrm{bands} M_\mathrm{band}$.
For the Northern Cross 5-band configuration this gives $M_\mathrm{hist} = 47.1$\,MB (exact, integer storage) and $4.6$\,MB (binary).
Without multi-rate scrunching the
exact-mode history buffer reaches 214\,MB, dominated by deep ring buffers at high DM (the dispersive
sweep at DM\,$= 3000$ spans ${\approx}42\,500$ native samples); multi-rate scrunching reduces
this by a factor of approximately $5\times$ in both modes.

Per-band resource figures are tabulated in Table~\ref{tab:resourcesXX}. Power estimates follow the model:
\begin{align}
  P_\mathrm{dyn}  &= N \cdot r \cdot f_S \cdot E_\mathrm{spike}, \label{eq:P_dyn}\\
  P_\mathrm{static} &= N \cdot f_S \cdot (E_\mathrm{leak} + E_\mathrm{SRAM}), \label{eq:P_static}\\
  P_\mathrm{hist}   &= W_\mathrm{buf} \cdot f_S \cdot E_\mathrm{mem}, \label{eq:P_hist}
\end{align}
where $N$ is the number of neurons in the band, $r$ is the mean fire rate, $f_S = 1/\Delta t_S$ is the scrunched sample rate, and
\begin{equation}
  W_\mathrm{buf} = (N_\mathrm{syn} + N_\mathrm{neu} + N_\mathrm{enc})\,\frac{\max(1,\,B_\mathrm{hist}/8)}{2}
  \label{eq:W_buf}
\end{equation}
is the aggregate number of ring-buffer accesses per timestep in 16-bit word equivalents. Here $N_\mathrm{syn}$ is the total number of SND synapses (each contributes one ring-buffer \emph{read} per timestep), $N_\mathrm{neu}$ is the number of SND neurons (each contributes one ring-buffer \emph{write} per timestep), $N_\mathrm{enc}$ is the number of encoder channels (one write per channel per timestep into the input spike buffer), and $B_\mathrm{hist}$ is the storage width per ring-buffer value ($1$\,bit for binary; $4$--$16$\,bit per level for the exact mode). The factor $\max(1, B_\mathrm{hist}/8)/2$ converts from bytes per access to 16-bit word equivalents. The aggregate ring-buffer bandwidth requirement is $\mathrm{BW} = W_\mathrm{buf} \times 2 \times f_S$\,bytes/s; this must not exceed the hardware memory bus capacity.

$E_\mathrm{mem}$ is $25$\,pJ per 16-bit word for LPDDR4 DRAM~\cite{jedec_lpddr4} or $0.25$\,pJ/word for on-chip SRAM. All SND power estimates use SpiNNaker~2 energy parameters: $E_\mathrm{spike} = 10$\,pJ, $E_\mathrm{leak} = 0.2$\,pJ/neuron/timestep, and $E_\mathrm{SRAM} = 0.5$\,pJ/neuron/timestep~\cite{hoppner2024spinnaker}. The history-buffer term $P_\mathrm{hist}$ depends only on memory technology (DRAM or on-chip SRAM) and applies equally to a Loihi~2 deployment.

Static and history-buffer power are purely model-derived. The static term evaluates to $P_\mathrm{static} = 0.42$\,mW for all modes. The history-buffer term evaluates to $0.61$\,mW (binary, on-chip SRAM) and $P_\mathrm{hist} = 62.5$\,mW (exact, DRAM, $5.0$\,GB/s bandwidth). The exact-mode DRAM bandwidth is within the SpiNNaker~2 LPDDR4 bandwidth limit of $25.6$\,GB/s~\cite{hoppner2024spinnaker}. The dynamic term $P_\mathrm{dyn}$ depends on per-level mean fire rates measured from the validation data; these values are reported in the power-consumption estimate.

The GPU reference power is measured from Heimdall running on an RTX~6000~Ada GPU, averaged over 30 Northern Cross filterbanks (130\,s each). With a mean active-window draw of $81.0 \pm 2.2$\,W and a duty cycle of $23.4\,\%$, the duty-cycle-adjusted real-time power is $18.94$\,W. This figure covers the full active-window draw of a production GPU system; the SND estimates above cover only the SND computation core, excluding board-level overhead. The power comparison is therefore between the dedispersion compute stages of each system, not between complete backends.

\subsection*{Operating-point selection}
\label{sec:opsweep}

The parameter sweep is conducted on a single representative Northern Cross filterbank (13 injections with fixed parameters spanning the expected FRB population). The recommended operating point is subsequently validated on the full 950-filterbank dataset. The binary-mode parameter space has five degrees of freedom: encoder threshold $\theta$, leaf threshold $K_L$, internal threshold $K_I$, detection S/N threshold $\Theta$, and matched-filter noise floor $\sigma_\mathrm{min}$. For the Northern Cross configuration, the sweep covers five values of $\theta \in \{0.75, 1.0, 1.25, 1.5, 2.0\}$\,$\sigma$ (encoder thresholds $\geq 2.5\sigma$ produce negligible sensitivity in binary mode on NC data and are excluded), eight $(K_L, K_I)$ pairs drawn from the viable operating window (Table~\ref{tab:binaryXX}) ($K_L \in \{1,2,3,4\}$, $K_I \in \{1,2\}$; $K_I \geq 3$ produces zero completeness across all injection configurations and is excluded), five detection thresholds $\Theta \in \{6, 7, 8, 9, 10\}$, and four noise-floor values $\sigma_\mathrm{min} \in \{0, 0.1, 0.15, 0.2\}$, for a total of $800$ binary-mode parameter combinations. The exact mode adds a further $25$ combinations (no noise-floor dimension; $\sigma_\mathrm{min}$ is not applied to non-binary planes).

The DM-time output plane is determined solely by $(\theta, K_L, K_I)$; varying $\Theta$ or $\sigma_\mathrm{min}$ affects only the downstream matched-filter and candidate-detection stages. The sweep is therefore organised as a two-stage procedure: a single encode-plus-SND run for each $(\theta, K_L, K_I)$ triple, followed by replaying the matched filter and candidate detection over the cached DM-time planes for all $(\Theta, \sigma_\mathrm{min})$ combinations.

Each result row records aggregate completeness and false-positive-per-injection count together with \emph{per-band completeness} $\{C_{S_1}, C_{S_2}, C_{S_4}, C_{S_8}, C_{S_{16}}\}$: the fraction of injected events whose DM falls within each scrunch band that are matched to a detected candidate. The per-band breakdown is essential because the cascade failure mechanism (the cascade noise analysis) affects high-S bands selectively; a configuration with high aggregate completeness may still produce zero completeness at $S = 8$ or $S = 16$. The recommended operating point is selected by maximising aggregate completeness subject to a false-positive budget, with the additional constraint that per-band completeness must be non-zero for all five bands. The exact operating point is selected by the same criterion with no noise-floor sweep.

\subsection*{Validation methodology}
\label{sec:validation}

Synthetic filterbanks are constructed from 30 real Northern Cross observation sessions, each contributing approximately 30--35 filterbank files of 130\,s duration, for a total of 950 filterbanks. Thirteen FRBs are injected at random times within each filterbank (minimum separation 5\,s to prevent overlap), yielding 12{,}350 injections in total. Each injection uses a Gaussian burst profile (the injection tool's ``SGB'' shape) with a single spectral component, no spectral index modulation, no sub-pulse drift, and no scattering tail. The injections were generated with a bespoke Northern Cross FRB injection tool; the resulting synthetic dataset is described in Ref.~\cite{camilleri2026}. Injection parameters are drawn independently: DM log-uniformly from $[20, 3000]$\,pc\,cm$^{-3}$; intrinsic width log-uniformly from $[0.5, 130]$\,ms; and peak S/N uniformly from $[3, 13]$. This injected S/N is the injection tool's native normalisation, in which a burst of value $\rho$ is placed with a single channel-sample peak amplitude of $\rho/\sqrt{N_f W}$. It is not the matched-filter S/N that a pipeline reports. On the matched candidates both pipelines recover a systematically higher detected S/N, with a median detected-to-injected ratio of $3.2$ for exact SND and $4.7$ for Heimdall. An injection labelled S/N $3$ is therefore recovered near detected S/N $10$ (SND) or $14$ (Heimdall), comfortably above the detection threshold $\Theta = 7$, which is why completeness stays high at the low end of the injected-S/N axis in Fig.~\ref{fig:validation}a. The log-uniform width prior spans the plausible observed range at $408$\,MHz and is a modelling choice. Because detection efficiency is strongly width-dependent, the aggregate completeness figures are population-dependent; the per-width breakdown (Fig.~\ref{fig:validation}b) should be used when comparing against a specific expected width distribution.

SND operating points depend on the tree activation mode and are listed in Supplementary Table S1. In brief, binary and exact modes use the delta encoder at $\theta = 1.5\sigma$ with detection threshold $\Theta = 7$; binary mode additionally applies quorum thresholds $K_L = 2$, $K_I = 2$ with $\sigma_\mathrm{min} = 0.20$. Heimdall is run with a DM search range of $0$--$3000$\,pc\,cm$^{-3}$, its standard DM tolerance $\tau = 1.05$, which the SND trial-DM grid matches, boxcar matched-filter widths in powers of two up to $512$ samples, its default detection threshold of $\mathrm{S/N} = 6$, and Heimdall's default RFI mitigation (baseline normalisation and zero-DM cleaning). Heimdall candidates below DM $= 10$\,pc\,cm$^{-3}$ are excluded to match the SND search range. Candidates from each pipeline are matched to injections by greedy one-to-one assignment in descending S/N order, accepting a match if $|\Delta \mathrm{DM}| \leq \max(10\,\mathrm{pc\,cm^{-3}},\, W_\mathrm{eff}/\kappa)$ and $|\Delta t| \leq \max(0.5\,W_\mathrm{eff},\, 25\,\mathrm{ms})$, where $W_\mathrm{eff}$ is the effective pulse width at the injection DM, with $W_\mathrm{eff} = \sqrt{W^2 + \tau_\mathrm{smear}^2 + \Delta t_\mathrm{samp}^2}$ and $\kappa = K_\mathrm{DM}(f_\mathrm{lo}^{-2} - f_\mathrm{hi}^{-2}) \approx 1.96$\,ms\,pc$^{-1}$\,cm$^3$ the full-band dispersive delay per unit DM. Detection completeness is computed as the fraction of injections matched by at least one candidate. The criterion is robust to chance coincidence, because a false candidate produces a spurious match only if it falls within the tolerance window of an injection. Given the measured false-positive densities (${\lesssim}500$ per file) and the $2990$\,pc\,cm$^{-3} \times 130$\,s search volume per file, the expected spurious-match rate is ${\lesssim}0.2\,\%$ of injections in every mode, far below the completeness differences between modes.

\backmatter

\bmhead{Acknowledgments}
The Northern Cross filterbank dataset used for validation and provided by Andrea Geminardi (IUSS Pavia; University of Trento; INAF, Osservatorio Astronomico di Cagliari). The author thanks the Northern Cross team at the Medicina Radio Observatory for access to the observing infrastructure.

\section*{Declarations}

\bmhead{Author contributions}
A.M.\ conceived and developed the Spiking Neural Dedispersion algorithm and multi-rate pipeline, designed and implemented the hardware resource model and power estimation framework, performed all numerical experiments, generated all figures, and wrote the manuscript in its entirety.

\bmhead{Data availability}
The dedispersion pipeline software is publicly available at \url{https://github.com/lessju/NeuroFRB}. Source data for Figure~5 and for Supplementary Figures~1 and~2 are provided as Supplementary Data. The synthetic Northern Cross filterbank dataset was produced with a third-party injection tool and cannot be redistributed under its licence; requests for access should be directed to the corresponding author (A.~Magro).

\bmhead{Competing interests}
The author declares no competing interests.

\bmhead{Funding Statement}
No external funding was received.

\bibliography{bibliography}

@article{lorimer2007,
    author = {D. R. Lorimer  and M. Bailes  and M. A. McLaughlin  and D. J. Narkevic  and F. Crawford },
    title = {A Bright Millisecond Radio Burst of Extragalactic Origin},
    journal = {Science},
    volume = {318},
    number = {5851},
    pages = {777-780},
    year = {2007},
    doi = {10.1126/science.1147532},
    URL = {https://www.science.org/doi/abs/10.1126/science.1147532},
    eprint = {https://www.science.org/doi/pdf/10.1126/science.1147532},
}

@article{Petroff2019,
  author    = {Petroff, E. and Hessels, J. W. T. and Lorimer, D. R.},
  title     = {Fast radio bursts},
  journal   = {The Astronomy and Astrophysics Review},
  year      = {2019},
  volume    = {27},
  number    = {1},
  pages     = {4},
  doi       = {10.1007/s00159-019-0116-6},
  url       = {https://doi.org/10.1007/s00159-019-0116-6},
  issn      = {1432-0754}
}

@article{petroff2022,
  author  = {Petroff, E. and Hessels, J. W. T. and Lorimer, D. R.},
  title   = {Fast radio bursts at the dawn of the 2020s},
  journal = {The Astronomy and Astrophysics Review},
  year    = {2022},
  volume  = {30},
  number  = {1},
  pages   = {2},
  doi     = {10.1007/s00159-022-00139-w},
  url     = {https://doi.org/10.1007/s00159-022-00139-w},
  issn    = {1432-0754}
}

@article{cordes2019,
  author    = {Cordes, J. M. and Chatterjee, S.},
  title     = {Fast Radio Bursts: An Extragalactic Enigma},
  journal   = {Annual Review of Astronomy and Astrophysics},
  volume    = {57},
  pages     = {417--465},
  year      = {2019},
  doi       = {10.1146/annurev-astro-091918-104501}
}

@article{chimefrb2021catalog,
doi = {10.3847/1538-4365/ac33ab},
url = {https://doi.org/10.3847/1538-4365/ac33ab},
year = {2021},
month = {dec},
publisher = {The American Astronomical Society},
volume = {257},
number = {2},
pages = {59},
author = {{The CHIME/FRB Collaboration} and Amiri, Mandana and Andersen, Bridget C. and Bandura, Kevin and Berger, Sabrina and Bhardwaj, Mohit and Boyce, Michelle M. and Boyle, P. J. and Brar, Charanjot and Breitman, Daniela and Cassanelli, Tomas and Chawla, Pragya and Chen, Tianyue and Cliche, J.-F. and Cook, Amanda and Cubranic, Davor and Curtin, Alice P. and Deng, Meiling and Dobbs, Matt and (Adam) Dong, Fengqiu and Eadie, Gwendolyn and Fandino, Mateus and Fonseca, Emmanuel and Gaensler, B. M. and Giri, Utkarsh and Good, Deborah C. and Halpern, Mark and Hill, Alex S. and Hinshaw, Gary and Josephy, Alexander and Kaczmarek, Jane F. and Kader, Zarif and Kania, Joseph W. and Kaspi, Victoria M. and Landecker, T. L. and Lang, Dustin and Leung, Calvin and Li, Dongzi and Lin, Hsiu-Hsien and Masui, Kiyoshi W. and Mckinven, Ryan and Mena-Parra, Juan and Merryfield, Marcus and Meyers, Bradley W. and Michilli, Daniele and Milutinovic, Nikola and Mirhosseini, Arash and Münchmeyer, Moritz and Naidu, Arun and Newburgh, Laura and Ng, Cherry and Patel, Chitrang and Pen, Ue-Li and Petroff, Emily and Pinsonneault-Marotte, Tristan and Pleunis, Ziggy and Rafiei-Ravandi, Masoud and Rahman, Mubdi and Ransom, Scott M. and Renard, Andre and Sanghavi, Pranav and Scholz, Paul and Shaw, J. Richard and Shin, Kaitlyn and Siegel, Seth R. and Sikora, Andrew E. and Singh, Saurabh and Smith, Kendrick M. and Stairs, Ingrid and Tan, Chia Min and Tendulkar, S. P. and Vanderlinde, Keith and Wang, Haochen and Wulf, Dallas and Zwaniga, A. V.},
title = {The First {CHIME/FRB} Fast Radio Burst Catalog},
journal = {The Astrophysical Journal Supplement Series},
}

@article{heintz2020,
doi = {10.3847/1538-4357/abb6fb},
url = {https://doi.org/10.3847/1538-4357/abb6fb},
year = {2020},
month = {nov},
publisher = {The American Astronomical Society},
volume = {903},
number = {2},
pages = {152},
author = {Heintz, Kasper E. and Prochaska, J. Xavier and Simha, Sunil and Platts, Emma and Fong, Wen-fai and Tejos, Nicolas and Ryder, Stuart D. and Aggerwal, Kshitij and Bhandari, Shivani and Day, Cherie K. and Deller, Adam T. and Kilpatrick, Charles D. and Law, Casey J. and Macquart, Jean-Pierre and Mannings, Alexandra and Marnoch, Lachlan J. and Sadler, Elaine M. and Shannon, Ryan M.},
title = {Host Galaxy Properties and Offset Distributions of Fast Radio Bursts: Implications for Their Progenitors},
journal = {The Astrophysical Journal},
}

@article{chimefrb2020magnetar,
  author  = {Andersen, B. C. and Bandura, K. M. and Bhardwaj, M. and Bij, A. and Boyce, M. M. and Boyle, P. J. and Brar, C. and Cassanelli, T. and Chawla, P. and Chen, T. and Cliche, J.-F. and Cook, A. and Cubranic, D. and Curtin, A. P. and Denman, N. T. and Dobbs, M. and Dong, F. Q. and Fandino, M. and Fonseca, E. and Gaensler, B. M. and Giri, U. and Good, D. C. and Halpern, M. and Hill, A. S. and Hinshaw, G. F. and Höfer, C. and Josephy, A. and Kania, J. W. and Kaspi, V. M. and Landecker, T. L. and Leung, C. and Li, D. Z. and Lin, H.-H. and Masui, K. W. and McKinven, R. and Mena-Parra, J. and Merryfield, M. and Meyers, B. W. and Michilli, D. and Milutinovic, N. and Mirhosseini, A. and Münchmeyer, M. and Naidu, A. and Newburgh, L. B. and Ng, C. and Patel, C. and Pen, U.-L. and Pinsonneault-Marotte, T. and Pleunis, Z. and Quine, B. M. and Rafiei-Ravandi, M. and Rahman, M. and Ransom, S. M. and Renard, A. and Sanghavi, P. and Scholz, P. and Shaw, J. R. and Shin, K. and Siegel, S. R. and Singh, S. and Smegal, R. J. and Smith, K. M. and Stairs, I. H. and Tan, C. M. and Tendulkar, S. P. and Tretyakov, I. and Vanderlinde, K. and Wang, H. and Wulf, D. and Zwaniga, A. V. and {The CHIME/FRB Collaboration}},
  title   = {A bright millisecond-duration radio burst from a Galactic magnetar},
  journal = {Nature},
  year    = {2020},
  volume  = {587},
  number  = {7832},
  pages   = {54--58}
}

@article{bochenek2020,
  author  = {Bochenek, C. D. and Ravi, V. and Belov, K. V. and Hallinan, G. and Kocz, J. and Kulkarni, S. R. and McKenna, D. L.},
  title   = {A fast radio burst associated with a Galactic magnetar},
  journal = {Nature},
  year    = {2020},
  volume  = {587},
  number  = {7832},
  pages   = {59--62},
  doi     = {10.1038/s41586-020-2872-x},
  url     = {https://doi.org/10.1038/s41586-020-2872-x},
  issn    = {1476-4687}
}

@article{macquart2020,
  author  = {Macquart, J.-P. and Prochaska, J. X. and McQuinn, M. and Bannister, K. W. and Bhandari, S. and Day, C. K. and Deller, A. T. and Ekers, R. D. and James, C. W. and Marnoch, L. and Os{\l}owski, S. and Phillips, C. and Ryder, S. D. and Scott, D. R. and Shannon, R. M. and Tejos, N.},
  title   = {A census of baryons in the Universe from localized fast radio bursts},
  journal = {Nature},
  year    = {2020},
  volume  = {581},
  number  = {7809},
  pages   = {391--395},
  doi     = {10.1038/s41586-020-2300-2},
  url     = {https://doi.org/10.1038/s41586-020-2300-2},
  issn    = {1476-4687}
}

@article{james2022,
    author = {James, C W and Prochaska, J X and Macquart, J-P and North-Hickey, F O and Bannister, K W and Dunning, A},
    title = {The z–DM distribution of fast radio bursts},
    journal = {Monthly Notices of the Royal Astronomical Society},
    volume = {509},
    number = {4},
    pages = {4775-4802},
    year = {2022},
    month = {02},
    issn = {0035-8711},
    doi = {10.1093/mnras/stab3051},
    url = {https://doi.org/10.1093/mnras/stab3051},
    eprint = {https://academic.oup.com/mnras/article-pdf/509/4/4775/41695367/stab3051.pdf},
}

@article{hagstotz2022,
  author    = {Hagstotz, S. and Reischke, R. and Lilow, R.},
  title     = {A new measurement of the {Hubble} constant using fast radio bursts},
  journal   = {Monthly Notices of the Royal Astronomical Society},
  volume    = {511},
  number    = {1},
  pages     = {662--667},
  year      = {2022},
  doi       = {10.1093/mnras/stac077}
}

@article{hessels2019,
doi = {10.3847/2041-8213/ab13ae},
url = {https://doi.org/10.3847/2041-8213/ab13ae},
year = {2019},
month = {may},
publisher = {The American Astronomical Society},
volume = {876},
number = {2},
pages = {L23},
author = {Hessels, J. W. T. and Spitler, L. G. and Seymour, A. D. and Cordes, J. M. and Michilli, D. and Lynch, R. S. and Gourdji, K. and Archibald, A. M. and Bassa, C. G. and Bower, G. C. and Chatterjee, S. and Connor, L. and Crawford, F. and Deneva, J. S. and Gajjar, V. and Kaspi, V. M. and Keimpema, A. and Law, C. J. and Marcote, B. and McLaughlin, M. A. and Paragi, Z. and Petroff, E. and Ransom, S. M. and Scholz, P. and Stappers, B. W. and Tendulkar, S. P.},
title = {{FRB} 121102 Bursts Show Complex Time–Frequency Structure},
journal = {The Astrophysical Journal Letters},
}

@article{nimmo2022,
  author  = {Nimmo, K. and Hessels, J. W. T. and Kirsten, F. and Keimpema, A. and Cordes, J. M. and Snelders, M. P. and Hewitt, D. M. and Karuppusamy, R. and Archibald, A. M. and Bezrukovs, V. and Bhardwaj, M. and Blaauw, R. and Buttaccio, S. T. and Cassanelli, T. and Conway, J. E. and Corongiu, A. and Feiler, R. and Fonseca, E. and Forss{\'e}n, O. and Gawro{\'n}ski, M. and Giroletti, M. and Kharinov, M. A. and Leung, C. and Lindqvist, M. and Maccaferri, G. and Marcote, B. and Masui, K. W. and McKinven, R. and Melnikov, A. and Michilli, D. and Mikhailov, A. G. and Ng, C. and Orbidans, A. and Ould-Boukattine, O. S. and Paragi, Z. and Pearlman, A. B. and Petroff, E. and Rahman, M. and Scholz, P. and Shin, K. and Smith, K. M. and Stairs, I. H. and Surcis, G. and Tendulkar, S. P. and Vlemmings, W. and Wang, N. and Yang, J. and Yuan, J. P.},
  title   = {Burst timescales and luminosities as links between young pulsars and fast radio bursts},
  journal = {Nature Astronomy},
  year    = {2022},
  volume  = {6},
  number  = {3},
  pages   = {393--401},
  doi     = {10.1038/s41550-021-01569-9},
  url     = {https://doi.org/10.1038/s41550-021-01569-9},
  issn    = {2397-3366}
}

@article{chimefrb2018system,
doi = {10.3847/1538-4357/aad188},
url = {https://doi.org/10.3847/1538-4357/aad188},
year = {2018},
month = {aug},
publisher = {The American Astronomical Society},
volume = {863},
number = {1},
pages = {48},
author = {{The CHIME/FRB Collaboration} and Amiri, M. and Bandura, K. and Berger, P. and Bhardwaj, M. and Boyce, M. M. and Boyle, P. J. and Brar, C. and Burhanpurkar, M. and Chawla, P. and Chowdhury, J. and Cliche, J.-F. and Cranmer, M. D. and Cubranic, D. and Deng, M. and Denman, N. and Dobbs, M. and Fandino, M. and Fonseca, E. and Gaensler, B. M. and Giri, U. and Gilbert, A. J. and Good, D. C. and Guliani, S. and Halpern, M. and Hinshaw, G. and Höfer, C. and Josephy, A. and Kaspi, V. M. and Landecker, T. L. and Lang, D. and Liao, H. and Masui, K. W. and Mena-Parra, J. and Naidu, A. and Newburgh, L. B. and Ng, C. and Patel, C. and Pen, U.-L. and Pinsonneault-Marotte, T. and Pleunis, Z. and Ravandi, M. Rafiei and Ransom, S. M. and Renard, A. and Scholz, P. and Sigurdson, K. and Siegel, S. R. and Smith, K. M. and Stairs, I. H. and Tendulkar, S. P. and Vanderlinde, K. and Wiebe, D. V.},
title = {The CHIME Fast Radio Burst Project: System Overview},
journal = {The Astrophysical Journal},
}

@techreport{vanderlinde2019chord,
  author     = {Vanderlinde, Keith and Liu, Adrian and Gaensler, Bryan and Bond, Dick and Hinshaw, Gary and Ng, Cherry and Chiang, Cynthia and Stairs, Ingrid and Brown, Jo-Anne and Sievers, Jonathan and Mena, Juan and Smith, Kendrick and Bandura, Kevin and Masui, Kiyoshi and Spekkens, Kristine and Belostotski, Leo and Dobbs, Matt and Turok, Neil and Boyle, Patrick and Rupen, Michael and Landecker, Tom and Pen, Ue-Li and Kaspi, Victoria},
  title     = {The {Canadian Hydrogen Observatory and Radio-transient Detector (CHORD)}},
  institution = {Canadian Long Range Plan for Astronomy and Astrophysics White Papers},
  year        = {2019},
  month       = {oct},
  type        = {White Paper},
  note        = {White paper identifier W028},
  doi         = {10.5281/zenodo.3765414},
  url         = {https://doi.org/10.5281/zenodo.3765414}
}

@article{lin2022burstt,
doi = {10.1088/1538-3873/ac8f71},
url = {https://doi.org/10.1088/1538-3873/ac8f71},
year = {2022},
month = {sep},
publisher = {The Astronomical Society of the Pacific},
volume = {134},
number = {1039},
pages = {094106},
author = {Lin, Hsiu-Hsien and Lin, Kai-yang and Li, Chao-Te and Tseng, Yao-Huan and Jiang, Homin and Wang, Jen-Hung and Cheng, Jen-Chieh and Pen, Ue-Li and Chen, Ming-Tang and Chen, Pisin and Chen, Yaocheng and Goto, Tomotsugu and Hashimoto, Tetsuya and Hwang, Yuh-Jing and King, Sun-Kun and Kubo, Derek and Kuo, Chung-Yun and Mills, Adam and Nam, Jiwoo and Oshiro, Peter and Shen, Chang-Shao and Tseng, Hsien-Chun and Wang, Shih-Hao and Wu, Vigo Feng-Shun and Bower, Geoffrey and Chang, Shu-Hao and Chen, Pai-An and Chen, Ying-Chih and Chiang, Yi-Kuan and Fedynitch, Anatoli and Gusinskaia, Nina and Ho, Simon C.-C. and Hsiao, Tiger Y.-Y. and Hu, Chin-Ping and Huang, Yau De and Jáuregui García, José Miguel and Kim, Seong Jin and Kuo, Cheng-Yu and Ling, Decmend Fang-Jie and On, Alvina Y. L. and Peterson, Jeffrey B. and R. Raquel, Bjorn Jasper and Su, Shih-Chieh and Uno, Yuri and Wu, Cossas K.-W. and Yamasaki, Shotaro and Zhu, Hong-Ming},
title = {{BURSTT}: Bustling Universe Radio Survey Telescope in Taiwan},
journal = {Publications of the Astronomical Society of the Pacific},
}

@misc{hallinan2019dsa,
  author       = {Hallinan, Gregg and Ravi, Vikram and Weinreb, Sander and Kulkarni, Shri and Kocz, Jon and McKenna, Dustin and Belov, Konstantin},
  title        = {The {DSA}-2000 -- A Radio Survey Camera},
  year         = {2019},
  eprint       = {1907.07648},
  archivePrefix= {arXiv},
  primaryClass = {astro-ph.IM},
  doi          = {10.48550/arXiv.1907.07648},
  url          = {https://arxiv.org/abs/1907.07648}
}

@inproceedings{braun2015ska,
  author    = {Braun, R. and Bourke, T. L. and Green, J. A. and Keane, E. and Wagg, J.},
  title     = {Advancing Astrophysics with the Square Kilometre Array},
  booktitle = {Proceedings of Advancing Astrophysics with the Square Kilometre Array},
  series    = {PoS(AASKA14)},
  address    = {Trieste, Italy},
  pages     = {174},
  year      = {2015},
  publisher = {SISSA Medialab},
  doi       = {10.22323/1.215.0174},
  url       = {https://pos.sissa.it/215/174}
}

@article{stappers2017meertrap, 
title={{MeerTRAP}: A pulsar and fast transients survey with {MeerKAT}},
volume={13}, 
DOI={10.1017/S1743921317009310}, 
number={S337}, 
journal={Proceedings of the International Astronomical Union}, author={Sanidas, S. and Caleb, M. and Driessen, L. and Morello, V. and Rajwade, K. and Stappers, B. W.}, 
year={2017}, 
pages={406–407}
}

@article{wangcraco,
title={The {CRAFT} coherent ({CRACO}) upgrade I: System description and results of the 110-ms radio transient pilot survey},
volume={42}, 
DOI={10.1017/pasa.2024.107}, 
journal={Publications of the Astronomical Society of Australia}, author={Wang, Z. and Bannister, K. W. and Gupta, V. and Deng, X. and Pilawa, M. and Tuthill, J. and Bunton, J. D. and Flynn, C. and Glowacki, M. and Jaini, A. and et al.}, 
year={2025}, 
pages={e005}
}

@article{naldi2017,
author = {Naldi, Giovanni and Mattana, Andrea and Pastore, Sandro and Alderighi, Monica and Zarb Adami, Kristian and Schillir\`{o}, Francesco and Aminaei, Amin and Baker, Jeremy and Belli, Carolina and Comoretto, Gianni and Chiarucci, Simone and Chiello, Riccardo and D’Angelo, Sergio and Dalle Mura, Gabriele and De Marco, Andrea and Halsall, Rob and Magro, Alessio and Monari, Jader and Roberts, Matt and Perini, Federico and Poloni, Marco and Pupillo, Giuseppe and Rusticelli, Simone and Schiaffino, Marco and Zaccaro, Emanuele},
title = {The Digital Signal Processing Platform for the Low Frequency Aperture Array: Preliminary Results on the Data Acquisition Unit},
journal = {Journal of Astronomical Instrumentation},
volume = {06},
number = {01},
pages = {1641014},
year = {2017},
doi = {10.1142/S2251171716410142},
URL = { https://doi.org/10.1142/S2251171716410142},
}

@inproceedings{naldi2025,
  author={Naldi, G. and Magro, A. and Fiori, F. and Mattana, A. and Perini, F. and Ragno, N. and Schiaffino, M. and Camilleri, H. and De Barro, A. and Bugeja, K. and Beduzzi, L. and Bernardi, G. and Bianchi, G. and Bruno, L. and De Luca, M. A. and Lizia, P. Di and Fiorentini, M. and Geminardi, A. and Massari, M. and Montaruli, M. F. and Orlati, A. and Pelliciari, D. and Pilia, M. and Poli, A. and Trudu, M.},
  booktitle={2025 URSI Asia-Pacific Radio Science Meeting (AP-RASC)}, 
  title={The design of a new digital signal processing system for the upgraded Northern Cross Radio Telescope}, 
  year={2025},
  volume={},
  number={},
  pages={1-4},
  doi={10.46620/URSIAPRASC25/ORKX6710}
}

@article{ pellicari2024,
	author = {{Pelliciari, D.} and {Bernardi, G.} and {Pilia, M.} and {Naldi, G.} and {Maccaferri, G.} and {Verrecchia, F.} and {Casentini, C.} and {Perri, M.} and {Kirsten, F.} and {Bianchi, G.} and {Bortolotti, C.} and {Bruno, L.} and {Dallacasa, D.} and {Esposito, P.} and {Geminardi, A.} and {Giarratana, S.} and {Giroletti, M.} and {Lulli, R.} and {Maccaferri, A.} and {Magro, A.} and {Mattana, A.} and {Perini, F.} and {Pupillo, G.} and {Roma, M.} and {Schiaffino, M.} and {Setti, G.} and {Tavani, M.} and {Trudu, M.} and {Zanichelli, A.}},
	title = {The Northern Cross Fast Radio Burst project - IV. Multi-wavelength study of the actively repeating FRB 20220912A},
	DOI= "10.1051/0004-6361/202450271",
	url= "https://doi.org/10.1051/0004-6361/202450271",
	journal = {A\&A},
	year = 2024,
	volume = 690,
	pages = "A219",
}

@article{sclocco2016,
  author  = {Sclocco, A. and van Leeuwen, J. and Bal, H. E. and van Nieuwpoort, R. V.},
  title   = {Real-time dedispersion for fast radio transient surveys, using auto tuning on many-core accelerators},
  journal = {Astronomy and Computing},
  volume  = {14},
  pages   = {1--7},
  year    = {2016},
  doi     = {10.1016/j.ascom.2016.01.001},
  url     = {https://doi.org/10.1016/j.ascom.2016.01.001}
}

@misc{pritchard2026impact,
      title={The Potential Impact of Neuromorphic Computing on Radio Telescope Observatories}, 
      author={Nicholas J. Pritchard and Richard Dodson and Andreas Wicenec},
      year={2026},
      eprint={2601.07130},
      archivePrefix={arXiv},
      primaryClass={astro-ph.IM},
      url={https://arxiv.org/abs/2601.07130}, 
}

@article{davies2018loihi,
  author={Davies, Mike and Srinivasa, Narayan and Lin, Tsung-Han and Chinya, Gautham and Cao, Yongqiang and Choday, Sri Harsha and Dimou, Georgios and Joshi, Prasad and Imam, Nabil and Jain, Shweta and Liao, Yuyun and Lin, Chit-Kwan and Lines, Andrew and Liu, Ruokun and Mathaikutty, Deepak and McCoy, Steven and Paul, Arnab and Tse, Jonathan and Venkataramanan, Guruguhanathan and Weng, Yi-Hsin and Wild, Andreas and Yang, Yoonseok and Wang, Hong},
  journal={IEEE Micro}, 
  title={{Loihi}: A Neuromorphic Manycore Processor with On-Chip Learning},
  year={2018},
  volume={38},
  number={1},
  pages={82-99},
  doi={10.1109/MM.2018.112130359}}

@article{davies2021loihi,
  author={Davies, Mike and Wild, Andreas and Orchard, Garrick and Sandamirskaya, Yulia and Guerra, Gabriel A. Fonseca and Joshi, Prasad and Plank, Philipp and Risbud, Sumedh R.},
  journal={Proceedings of the IEEE}, 
  title={Advancing Neuromorphic Computing With {Loihi}: A Survey of Results and Outlook},
  year={2021},
  volume={109},
  number={5},
  pages={911-934},
  doi={10.1109/JPROC.2021.3067593}}

@ARTICLE{painkras2012spinnaker,
  author  = {Painkras, Eustace and Plana, Luis A. and Garside, Jim and Temple, Steve and Galluppi, Francesco and Patterson, Cameron and Lester, David R. and Brown, Andrew D. and Furber, Steve B.},
  journal = {IEEE Journal of Solid-State Circuits},
  title   = {{SpiNNaker}: A 1-{W} 18-Core System-on-Chip for Massively-Parallel Neural Network Simulation},
  year    = {2013},
  volume  = {48},
  number  = {8},
  pages   = {1943--1953},
  doi     = {10.1109/JSSC.2013.2259038}
}

@misc{mayr2019spinnaker,
      title={SpiNNaker 2: A 10 Million Core Processor System for Brain Simulation and Machine Learning}, 
      author={Christian Mayr and Sebastian Hoeppner and Steve Furber},
      year={2019},
      eprint={1911.02385},
      archivePrefix={arXiv},
      primaryClass={cs.ET},
      url={https://arxiv.org/abs/1911.02385}, 
}

@misc{hoppner2024spinnaker,
      title={The SpiNNaker 2 Processing Element Architecture for Hybrid Digital Neuromorphic Computing},
      author={Sebastian Höppner and Yexin Yan and Andreas Dixius and Stefan Scholze and Johannes Partzsch and Marco Stolba and Florian Kelber and Bernhard Vogginger and Felix Neumärker and Georg Ellguth and Stephan Hartmann and Stefan Schiefer and Thomas Hocker and Dennis Walter and Genting Liu and Jim Garside and Steve Furber and Christian Mayr},
      year={2022},
      eprint={2103.08392},
      archivePrefix={arXiv},
      primaryClass={cs.AR},
      url={https://arxiv.org/abs/2103.08392},
}

@misc{gonzalez2024spinnaker2,
      title={SpiNNaker2: A Large-Scale Neuromorphic System for Event-Based and Asynchronous Machine Learning},
      author={Hector A. Gonzalez and Jiaxin Huang and Florian Kelber and Khaleelulla Khan Nazeer and Tim Langer and Chen Liu and Matthias Lohrmann and Amirhossein Rostami and Mark Schöne and Bernhard Vogginger and Timo C. Wunderlich and Yexin Yan and Mahmoud Akl and Christian Mayr},
      year={2024},
      eprint={2401.04491},
      archivePrefix={arXiv},
      primaryClass={cs.AR},
      url={https://arxiv.org/abs/2401.04491},
      note={Workshop on Machine Learning with New Compute Paradigms, NeurIPS 2023.},
}

@misc{intel_halapoint_2024,
  author       = {{Intel Corporation}},
  title        = {Hala Point: Large-Scale Neuromorphic System},
  year         = {2024},
  howpublished = {Intel Newsroom Technical Announcement},
  url          = {https://newsroom.intel.com/artificial-intelligence/intel-builds-worlds-largest-neuromorphic-system-to-enable-more-sustainable-ai/},
  note         = {Accessed: 2026-05-14}
}

@article{pritchard2024rfi,
  author  = {Pritchard, Nicholas J. and Wicenec, Andreas and Bennamoun, Mohammed and Dodson, Richard},
  title   = {Spiking neural networks for radio frequency interference detection in radio astronomy},
  journal = {Communications Physics},
  year    = {2025},
  volume  = {8},
  number  = {1},
  pages   = {517},
  doi     = {10.1038/s42005-025-02420-7},
  url     = {https://doi.org/10.1038/s42005-025-02420-7},
  issn    = {2399-3650}
}

@misc{pritchard2025rfi,
      title={Neuromorphic Astronomy: An End-to-End SNN Pipeline for RFI Detection Hardware}, 
      author={Nicholas J. Pritchard and Andreas Wicenec and Richard Dodson and Mohammed Bennamoun and Dylan R. Muir},
      year={2025},
      eprint={2511.16060},
      archivePrefix={arXiv},
      primaryClass={cs.NE},
      url={https://arxiv.org/abs/2511.16060}, 
}

@article{lopex2022,
   title={Time-Coded Spiking Fourier Transform in Neuromorphic Hardware},
   volume={71},
   ISSN={2326-3814},
   url={http://dx.doi.org/10.1109/TC.2022.3162708},
   DOI={10.1109/tc.2022.3162708},
   number={11},
   journal={IEEE Transactions on Computers},
   publisher={Institute of Electrical and Electronics Engineers (IEEE)},
   author={Lopez-Randulfe, Javier and Reeb, Nico and Karimi, Negin and Liu, Chen and Gonzalez, Hector A. and Dietrich, Robin and Vogginger, Bernhard and Mayr, Christian and Knoll, Alois},
   year={2022},
   month=Nov, pages={2792–2802} }

@inproceedings{aimone2019,
author = {Aimone, James B. and Ho, Yang and Parekh, Ojas and Phillips, Cynthia A. and Pinar, Ali and Severa, William and Wang, Yipu},
title = {Provable Advantages for Graph Algorithms in Spiking Neural Networks},
year = {2021},
isbn = {9781450380706},
publisher = {Association for Computing Machinery},
address = {New York, NY, USA},
url = {https://doi.org/10.1145/3409964.3461813},
doi = {10.1145/3409964.3461813},
booktitle = {Proceedings of the 33rd ACM Symposium on Parallelism in Algorithms and Architectures},
pages = {35–47},
numpages = {13},
location = {Virtual Event, USA},
series = {SPAA '21}
}

@book{lorimer2004handbook,
  author    = {Lorimer, D. R. and Kramer, M.},
  title     = {Handbook of Pulsar Astronomy},
  publisher = {Cambridge University Press},
  year      = {2004},
  series    = {Cambridge Observing Handbooks for Research Astronomers},
  isbn      = {9780521828239},
  address = {Cambridge, UK},
  url       = {https://ui.adsabs.harvard.edu/abs/2004hpa..book.....L}
}

@article{eatough2009,
  author  = {Eatough, R. P. and Keane, E. F. and Lyne, A. G.},
  title   = {An interference removal technique for radio pulsar searches},
  journal = {Monthly Notices of the Royal Astronomical Society},
  year    = {2009},
  volume  = {395},
  number  = {1},
  pages   = {410--415},
  doi     = {10.1111/j.1365-2966.2009.14524.x},
  url     = {https://doi.org/10.1111/j.1365-2966.2009.14524.x}
}

@article{cordes2003transients,
  author  = {Cordes, J. M. and McLaughlin, M. A.},
  title   = {Searches for Fast Radio Transients},
  journal = {The Astrophysical Journal},
  year    = {2003},
  volume  = {596},
  number  = {2},
  pages   = {1142--1154},
  doi     = {10.1086/378231},
  url     = {https://doi.org/10.1086/378231}
}

@misc{intel_loihi2_brief,
  author       = {{Intel Corporation}},
  title        = {Taking Neuromorphic Computing to the Next Level with {Loihi~2}},
  year         = {2021},
  howpublished = {Intel Technology Brief},
  url          = {https://www.intel.com/content/dam/www/central-libraries/us/en/documents/neuromorphic-computing-loihi-2-brief.pdf},
  note         = {Accessed: 2026-06-03.}
}

@article{taylor1974,
  author    = {Taylor, J. H.},
  title     = {A Sensitive Method for Detecting Dispersed Radio Emission},
  journal   = {Astronomy and Astrophysics Supplement Series},
  volume    = {15},
  pages     = {367--375},
  year      = {1974}
}

@article{zackay2017fdmt,
doi = {10.3847/1538-4357/835/1/11},
url = {https://doi.org/10.3847/1538-4357/835/1/11},
year = {2017},
month = {jan},
publisher = {The American Astronomical Society},
volume = {835},
number = {1},
pages = {11},
author = {Zackay, Barak and Ofek, Eran O.},
title = {An Accurate and Efficient algorithm for Detection of Radio Bursts with an Unknown Dispersion Measure, for Single-Dish Telescopes and Interferometers},
journal = {The Astrophysical Journal},
}

@article{barsdell2012,
  author    = {Barsdell, B. R. and Bailes, M. and Barnes, D. G. and Fluke, C. J.},
  title     = {Accelerating Incoherent Dedispersion},
  journal   = {Monthly Notices of the Royal Astronomical Society},
  volume    = {422},
  number    = {1},
  pages     = {379--392},
  year      = {2012},
  doi       = {10.1111/j.1365-2966.2012.20622.x}
}

@article{Tingay2013MWA,
  author  = {Tingay, S. J. and Goeke, R. and Bowman, J. D. and Emrich, D. and Ord, S. M. and Mitchell, D. A. and Morales, M. F. and Booler, T. and Crosse, B. and Wayth, R. B. and Lonsdale, C. J. and Tremblay, S. E. and Pallot, D. and Colegate, T. and Wicenec, A. and Kudryavtseva, N. and others},
  title   = {The Murchison Widefield Array: The Square Kilometre Array Precursor at Low Radio Frequencies},
  journal = {Publications of the Astronomical Society of Australia},
  year    = {2013},
  volume  = {30},
  pages   = {e007},
  doi     = {10.1017/pasa.2012.007},
  url     = {https://doi.org/10.1017/pasa.2012.007}
}

@inproceedings{Barsdell2012Heimdall,
  author    = {Barsdell, B. R. and Bailes, M. and Barnes, D. G. and Fluke, C. J.},
  title     = {Accelerating incoherent dedispersion},
  booktitle = {Proceedings of the International Conference on High Performance Computing \& Simulation (HPCS)},
  year      = {2012},
  pages     = {96--103},
  doi       = {10.1109/HPCSim.2012.6266923},
  url       = {https://doi.org/10.1109/HPCSim.2012.6266923}
}

@article{Selina2018ngVLA,
  author  = {Selina, R. J. and Murphy, E. J. and McKinnon, M. and Beasley, A. and Butler, B. and Carilli, C. and Clark, B. and Erickson, A. and Grammer, W. and Jackson, J. and Kent, B. and Mason, B. and Morgan, M. and Ojeda, O. and Shillue, W. and Sturgis, S. and Urbain, D.},
  title   = {The Next Generation Very Large Array: A Technical Overview},
  journal = {arXiv e-prints},
  year    = {2018},
  eprint  = {1806.08405},
  archivePrefix = {arXiv},
  primaryClass   = {astro-ph.IM},
  url     = {https://arxiv.org/abs/1806.08405}
}

@article{Hotan2021ASKAP,
  author  = {Hotan, A. W. and Bunton, J. D. and Chippendale, A. P. and Whiting, M. and Tuthill, J. and Moss, V. A. and McConnell, D. and Amy, S. W. and Huynh, M. T. and Allison, J. R. and Anderson, C. S. and Bannister, K. W. and Bastholm, E. and Beresford, R. and Bock, D. C.-J. and Bolton, R. and Chapman, J. M. and Chow, K. and Collier, J. D. and Cooray, F. R. and Cornwell, T. J. and Diamond, P. J. and Edwards, P. G. and Feain, I. J. and Franzen, T. M. O. and George, D. and Gupta, N. and Hampson, G. A. and Harvey-Smith, L. and Hayman, D. B. and Heywood, I. and Jacka, C. and Jackson, C. A. and Jackson, S. and Jeganathan, K. and Johnston, S. and Kesteven, M. and Kleiner, D. and Koribalski, B. S. and Lee-Waddell, K. and Lenc, E. and Lensson, E. S. and Mackay, S. and Mahony, E. K. and McClure-Griffiths, N. M. and McConigley, R. and Mirtschin, P. and Ng, A. K. and Norris, R. P. and Pearce, S. E. and Phillips, C. and Pilawa, M. A. and Raja, W. and Reynolds, J. E. and Roberts, P. and Roxby, D. N. and Sadler, E. M. and Shields, M. and Schinckel, A. E. T. and Serra, P. and Shaw, R. D. and Sweetnam, T. and Troup, E. R. and Tzioumis, A. and Voronkov, M. A. and Westmeier, T.},
  title   = {Australian Square Kilometre Array Pathfinder: I. System Description},
  journal = {Publications of the Astronomical Society of Australia},
  year    = {2021},
  volume  = {38},
  pages   = {e009},
  doi     = {10.1017/pasa.2021.1},
  url     = {https://doi.org/10.1017/pasa.2021.1}
}

@misc{isik2024loihi2,
  author        = {Isik, Murat and Tiwari, Karn and Eryilmaz, Muhammed Burak and Dikmen, I. Can},
  title         = {Accelerating Sensor Fusion in Neuromorphic Computing: A Case Study on {Loihi-2}},
  year          = {2024},
  eprint        = {2408.16096},
  archivePrefix = {arXiv},
  primaryClass  = {cs.NE},
  url           = {https://arxiv.org/abs/2408.16096}
}

@misc{jedec_lpddr4,
  author       = {{JEDEC Solid State Technology Association}},
  title        = {{LPDDR4: Low Power Double Data Rate 4 SDRAM Standard (JESD209-4D)}},
  year         = {2020},
  howpublished = {JEDEC Standard},
  url          = {https://www.jedec.org/standards-documents/docs/jesd209-4d},
}

@article{camilleri2026,
  author        = {Camilleri, H. and Magro, A. and Geminardi, A. and Naldi, G. and Bernardi, G. and Bruno, L. and Cesare, V. and Fiori, F. and Pelliciari, D. and Pilia, M. and Trudu, M.},
  title         = {Optimising the {FRB} Search Pipeline for the {Northern Cross} Radio Telescope},
  year          = {2026},
  eprint        = {2603.16345},
  archivePrefix = {arXiv},
  primaryClass  = {astro-ph.IM}
}

@inproceedings{naldi2026gass,
  author    = {Naldi, G. and others},
  title     = {Paving the Way for the Next Generation {Northern Cross} Radio Telescope},
  booktitle = {2026 URSI General Assembly and Scientific Symposium (GASS)},
  year      = {2026},
  address   = {Krak\'ow, Poland},
  note      = {Session J01}
}

@phdthesis{ransom2001presto,
  author = {Ransom, Scott M.},
  title  = {New search techniques for binary pulsars},
  school = {Harvard University},
  year   = {2001},
  note   = {PRESTO: PulsaR Exploration and Search TOolkit, \url{https://github.com/scottransom/presto}}
}

@article{mead1990,
  author  = {Mead, Carver},
  journal = {Proceedings of the IEEE},
  title   = {Neuromorphic electronic systems},
  year    = {1990},
  volume  = {78},
  number  = {10},
  pages   = {1629-1636},
  doi     = {10.1109/5.58356}}

@article{schuman2022,
  author  = {Schuman, Catherine D. and Kulkarni, Shruti R. and Parsa, Maryam and Mitchell, J. Parker and Date, Prasanna and Kay, Bill},
  journal = {Nature Computational Science},
  title   = {Opportunities for neuromorphic computing algorithms and applications},
  year    = {2022},
  volume  = {2},
  number  = {1},
  pages   = {10-19},
  doi     = {10.1038/s43588-021-00184-y}}

\end{document}


\begin{center}
{\Large\bfseries Supplementary Information}\\[1em]
{\large Neuromorphic Dedispersion for Fast Radio Burst Detection}
\end{center}

\section*{Glossary of hardware acronyms}

\noindent
\begin{description}
\item[SRAM] Static random-access memory: fast on-chip memory holding neuron state and, in binary mode, the dedispersion history buffer.
\item[DRAM] Dynamic random-access memory: higher-capacity off-chip memory used for the history buffer in the exact mode.
\item[LPDDR4] Low-Power Double Data Rate 4: the off-chip DRAM standard assumed for the external memory interface (25.6\,GB/s on SpiNNaker~2).
\item[TDP] Thermal design power: the sustained power envelope a chip is designed to dissipate.
\item[GbE] Gigabit Ethernet: the board-level network interface used for spike/data ingress.
\item[FD-SOI] Fully-depleted silicon-on-insulator: the 22\,nm fabrication process of SpiNNaker~2.
\item[SPIF] SpiNNaker Peripheral Interface: the spike-injection peripheral that streams external events into the SpiNNaker~2 fabric.
\item[Oheo Gulch] The development/test board hosting the Loihi~2 chip together with its external LPDDR4 memory.
\end{description}

\clearpage

\section*{Supplementary Tables}

\noindent This file contains supplementary tables associated with the main manuscript. Tables are numbered independently from the main text as Supplementary Tables~\ref{tab:supp-ncparams}--\ref{tab:supp-platforms}.

\begin{table}[htbp]
\centering
\small
\caption{Northern Cross instrument and SND pipeline parameters used throughout this work.}
\label{tab:supp-ncparams}
\begin{tabular}{ll}
\toprule
Parameter & Value \\
\midrule
\multicolumn{2}{@{}l}{\textit{Instrument}} \\
Centre frequency $f_c$           & 408\,MHz \\
Bandwidth                        & 16\,MHz \\
Number of channels $N_f$         & 1024 \\
Per-channel bandwidth $\Delta f$ & 15.6\,kHz \\
Sampling interval $\Delta t$     & 0.138\,ms \\
\midrule
\multicolumn{2}{@{}l}{\textit{DM search}} \\
DM range                         & $[10, 3000]$\,pc\,cm$^{-3}$ \\
Number of DM trials              & 3435 (Heimdall-spaced, $\tau = 1.05$) \\
\midrule
\multicolumn{2}{@{}l}{\textit{SND tree}} \\
Cluster size $C$                 & 4 \\
Branching factor $B$             & 8 \\
Root branching factor $B_\mathrm{root}$ & 4 \\
Number of tree levels            & 4 (leaf + L1 + L2 + root) \\
\midrule
\multicolumn{2}{@{}l}{\textit{Multi-rate bands}} \\
$S = 1$  & DM $\in [10, 289]$\,pc\,cm$^{-3}$,   $\Delta t_S = 0.138$\,ms \\
$S = 2$  & DM $\in [290, 578]$\,pc\,cm$^{-3}$,  $\Delta t_S = 0.276$\,ms \\
$S = 4$  & DM $\in [579, 1156]$\,pc\,cm$^{-3}$, $\Delta t_S = 0.552$\,ms \\
$S = 8$  & DM $\in [1158, 2311]$\,pc\,cm$^{-3}$, $\Delta t_S = 1.104$\,ms \\
$S = 16$ & DM $\in [2313, 3000]$\,pc\,cm$^{-3}$, $\Delta t_S = 2.208$\,ms \\
\midrule
\multicolumn{2}{@{}l}{\textit{Operating points by mode}} \\
\multicolumn{2}{@{}l}{\textit{Binary}} \\
Encoder                           & delta, $\theta = 1.5\,\sigma$ \\
Tree thresholds                   & $K_L = 2$, $K_I = 2$ \\
Noise floor $\sigma_\mathrm{min}$ & 0.20 \\
Detection threshold $\Theta$      & 7.0 \\
Zero-DM threshold $\gamma_\mathrm{zdm}$ & 5.0 \\
\midrule
\multicolumn{2}{@{}l}{\textit{Exact}} \\
Encoder                           & delta, $\theta = 1.5\,\sigma$ \\
Tree thresholds                   & $K_L = 0$, $K_I = 0$ (accumulate, no quorum) \\
Detection threshold $\Theta$      & 7.0 \\
Zero-DM threshold $\gamma_\mathrm{zdm}$ & 5.0 \\
\bottomrule
\end{tabular}
\end{table}

\begin{table}[htbp]
\centering
\small
\caption{Estimated power budget for the SND multi-rate pipeline on the Northern Cross instrument. Dynamic terms are from measured firing rates; static and history-buffer terms are modelled as described in the Methods section of the main manuscript. SND estimates use SpiNNaker~2 energy parameters ($E_\mathrm{spike} = 10$\,pJ, $E_\mathrm{leak} = 0.2$\,pJ/neuron/timestep, $E_\mathrm{SRAM} = 0.5$\,pJ/neuron/timestep); the history-buffer term applies equally to Loihi~2. Binary operating point: $K_L{=}2$, $K_I{=}2$. All figures are algorithmic projections from published chip energy parameters, excluding board-level baseline power; hardware validation is required. The exact column stores the per-level partial sums at their natural integer widths ($4$--$16$\,bit for this tree); naive float32 storage of the same values is bit-identical but raises the history-buffer term to $243.2$\,mW (grand total $248.4$\,mW). Bold indicates the grand totals.}
\label{tab:supp-power}
\begin{tabular}{lrr}
\toprule
Stage & Exact (mW) & Binary (mW) \\
\midrule
Encoder (dynamic)         &   0.010 &   0.011 \\
SND leaf L0 (dynamic)     &   0.201 &   0.027 \\
SND internal L1--L2 (dyn) &   4.379 &   0.121 \\
SND root L3 (dynamic)     &   0.148 &   0.000 \\
\midrule
Total dynamic             &   4.738 &   0.160 \\
Leakage                   &   0.119 &   0.119 \\
State SRAM access         &   0.298 &   0.298 \\
\midrule
History buffer (DRAM/SRAM) & 62.5 & 0.61 \\
\midrule
\textbf{Grand total}      & \textbf{67.6} & \textbf{1.19} \\
\bottomrule
\end{tabular}
\end{table}

\begin{table}[htbp]
\centering
\small
\caption{Key specifications of the neuromorphic hardware platforms used for pipeline mapping in this work. The per-spike energy for Loihi~2 is taken from a sensor-fusion benchmark; no direct packet-energy equivalent to the SpiNNaker~2 measurement is available in the literature. Memory-access power estimates assume LPDDR4 DRAM at $25$\,pJ/16-bit word for off-chip accesses, or on-chip SRAM at $0.25$\,pJ/16-bit word.}
\label{tab:supp-platforms}
\small
\setlength{\tabcolsep}{5pt}
\begin{tabular}{@{}p{3.7cm} p{5.7cm} p{5.7cm}@{}}
\toprule
Specification & SpiNNaker~2 & Loihi~2 \\
\midrule
Fabrication process    & GF 22\,nm FD-SOI & Intel~4 (7\,nm EUV) \\
Neurons per chip       & $\approx$152\,000 & $\approx$1\,000\,000 \\
On-chip SRAM           & 19\,MB (shared pool) & $\approx$24\,MB ($\approx$192\,KB per core across 128 cores) \\
External memory        & LPDDR4, 25.6\,GB/s & LPDDR4 (Oheo Gulch board) \\
SRAM access energy     & $\approx$0.25\,pJ/16-bit word & $\approx$0.25\,pJ/16-bit word \\
DRAM access energy     & $\approx$25\,pJ/16-bit word (LPDDR4) & $\approx$25\,pJ/16-bit word (LPDDR4) \\
Minimum timestep       & 1\,ms (typical) & 1\,ms (typical) \\
Spike payload          & 32-bit int or float32 & 32-bit integer only \\
Per-spike energy       & $\approx$10\,pJ & $\approx$10--12\,pJ \\
Max synaptic delay     & Routing schedule (no fixed cap) & 63 timesteps \\
SND exact mode     & int32 accumulation (native) & int32 accumulation (native) \\
\bottomrule
\end{tabular}
\end{table}

\clearpage

\section*{Supplementary Figures}

\begin{figure}[htbp]
\centering
\includegraphics[width=0.95\textwidth]{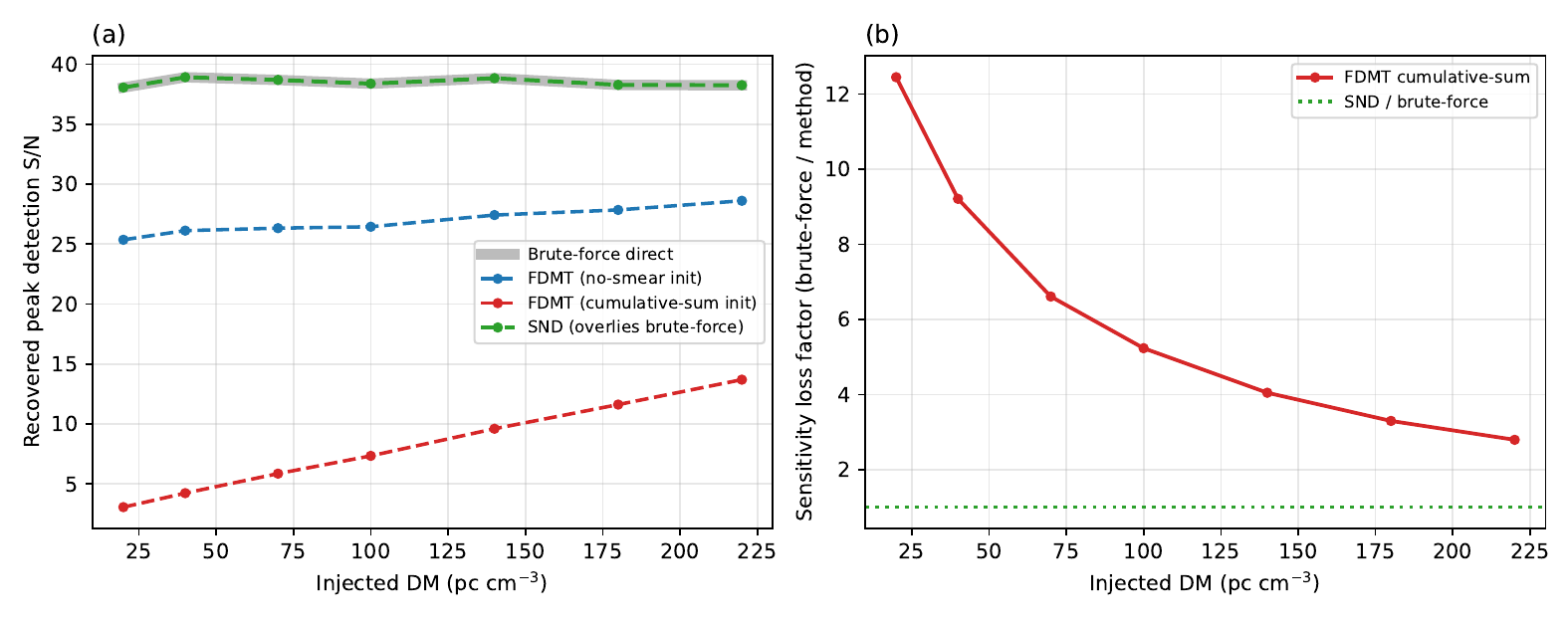}
\caption{Standard FDMT initialisation loses detection sensitivity on spike-encoded inputs, whereas SND matches brute-force dedispersion. A dispersed pulse (injected signal-to-noise ratio $60$, intrinsic width $1$\,ms, Northern Cross geometry with $512$ channels) is delta-encoded into a binary spike raster and dedispersed by four methods. \textbf{(a)}~Recovered peak signal-to-noise ratio for brute-force direct dedispersion (thick grey line), SND (green dashed line, overlying the brute-force curve), FDMT with the no-smear initialisation (blue dashed line), and FDMT with the standard cumulative-sum initialisation (red dashed line). \textbf{(b)}~Sensitivity loss factor, defined as the brute-force value divided by the method value, for the cumulative-sum FDMT (red solid line with circles); the loss grows from about three at high dispersion measure to about twelve at low dispersion measure. SND (green dotted line) is lossless. Points are means over three independent noise realisations.}
\label{fig:supp-fdmt}
\end{figure}

\clearpage

\begin{figure}[t]
\centering
\includegraphics[width=0.95\textwidth]{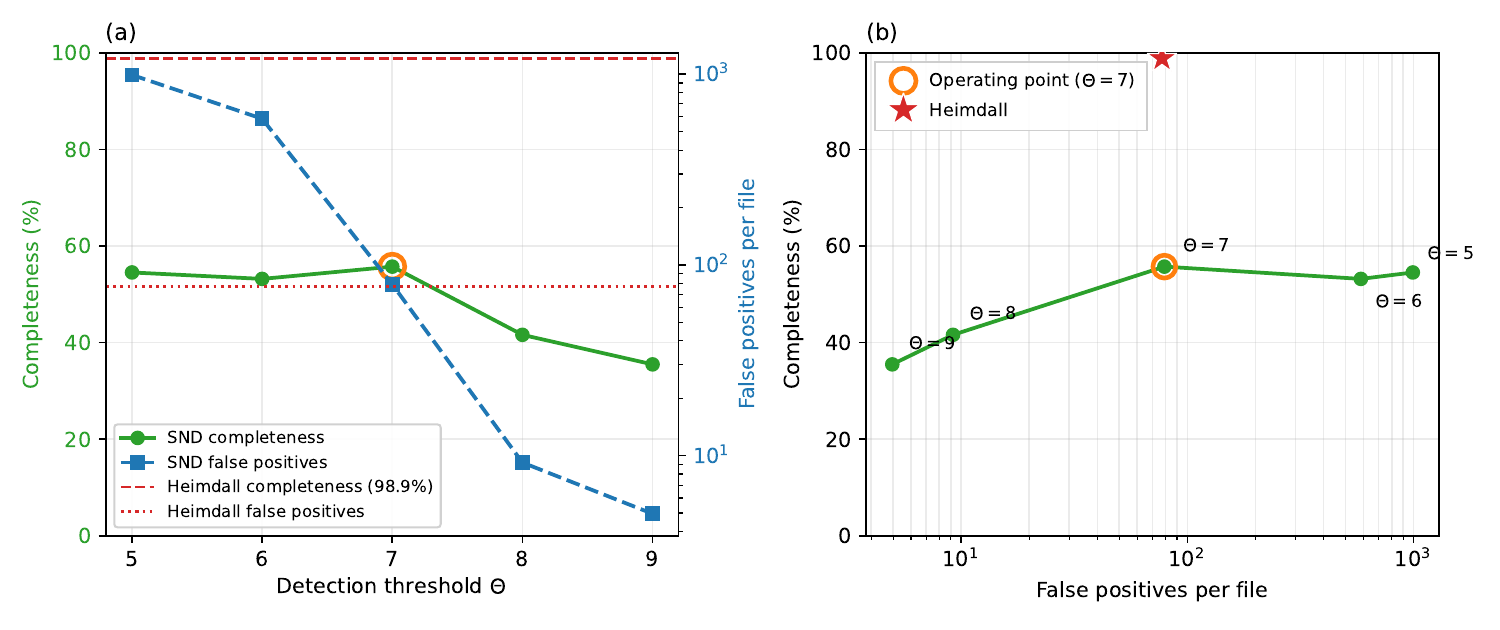}
\caption{Binary-mode completeness and false-positive trade-off against detection threshold, compared with the Heimdall reference. The binary pipeline is run at five detection thresholds on the 950-file Northern Cross validation set (12{,}350 injections). \textbf{(a)}~Detection completeness (green circles, left axis) and mean false positives per file (blue squares, right axis, logarithmic scale) against detection threshold; the dashed red line marks the Heimdall completeness. \textbf{(b)}~Completeness against mean false positives per file for the binary pipeline (green line with circles), with each detection threshold labelled, and the Heimdall operating point (red star). False positives fall steeply and monotonically with threshold while completeness stays near its maximum up to a threshold of seven and then declines; the threshold of seven thus has both higher completeness and far fewer false positives than the lower thresholds. At a threshold of seven the binary pipeline reaches $55.7\%$ completeness at $79$ false positives per file; at the matched false-positive rate Heimdall reaches $98.9\%$, so the binary-mode completeness gap reflects the one-bit quorum accumulation and is recovered by the exact mode.}
\label{fig:supp-threshold}
\end{figure}

\section*{Supplementary Data}

\noindent Source data for the figures are provided as comma-separated-value files. \texttt{Figure5a}, \texttt{Figure5b} and \texttt{Figure5c} give detection completeness against injected signal-to-noise ratio, pulse width and dispersion measure (per-bin completeness, binomial standard error, and bin counts for each mode and the Heimdall reference). \texttt{Figure5e} gives the dispersion-measure recovery residual profile (median and 16th--84th percentile per bin). \texttt{Figure5f} gives the false-positive count per observation file for each pipeline. \texttt{SupplementaryFigure1} gives the recovered peak signal-to-noise ratio for each dedisperser against dispersion measure, and \texttt{SupplementaryFigure2} gives binary-mode completeness and false positives per file against detection threshold.

%
%